\documentclass[fleqn,usenatbib]{mnras}

\usepackage{newtxtext,newtxmath}

\usepackage[T1]{fontenc}
\usepackage{ae,aecompl}

\usepackage{gensymb}

\usepackage{graphicx}	
\usepackage{amsmath}	
\usepackage{amssymb}	

\usepackage{animate}[2017/05/18]
\usepackage{algpseudocode}
\usepackage{algorithmicx}
\usepackage{algorithm}

\DeclareMathOperator{\prox}{prox}

\DeclareMathOperator{\soft}{\op{S}}
\DeclareMathOperator{\identity}{\op{I}}

\DeclareMathOperator{\proj}{\op{P}}
\DeclareMathOperator*{\argmin}{argmin}

\newcommand{\ds}{\displaystyle}
\newcommand{\hquad}{\;\:}

\newcommand{\bs}{\boldsymbol}
\newcommand{\bm}[1]{\boldsymbol{\mathsf{#1}}}
\newcommand{\bb}{\mathbb}
\newcommand{\mc}{\mathcal}
\newcommand{\diff}{\mathrm{d}\hspace{-0.1ex}}

\newcommand{\ac}[1]{\uppercase{#1}}
\newcommand{\alg}[1]{\textsc{#1}}
\newcommand{\op}[1]{\boldsymbol{\mc{#1}}}

\newcommand{\bcs}{\color{red}}

\algblock[Block]{Block}{EndBlock}
\algblockdefx[Block]{Block}{EndBlock}%
[1]{{#1}}%
[1]{{#1}}

\newcommand{\Given}[1]{\State{\bf given} {#1}}
\newcommand{\Output}[1]{\State{\bf output} {#1}}

\newcommand{\RepeatFor}[1]{\Repeat {\bf~for} {#1}}

\newcommand{\ParFor}[1]{\Block{{#1} \bf do~in~parallel}}
\newcommand{\EndParFor}{\EndBlock{\bf end}}

\newcommand{\bc}{\color{black}}
\newcommand{\ec}{\color{black}}

\title[Cyg A revisited]{Cygnus A super-resolved via convex optimisation from VLA data}

\author[A. Dabbech et al.]{
A. Dabbech,$^{1}$\thanks{E-mail: a.dabbech@hw.ac.uk}
A. Onose,$^{1}$
A. Abdulaziz,$^{1}$
R.A. Perley,$^{2}$
O.M. Smirnov$^{3,4}$
and Y. Wiaux$^{1}$
\\
$^{1}$Institute of Sensors, Signals and Systems, Heriot-Watt University, Edinburgh EH14 4AS, UK\\
$^{2}$National Radio Astronomy Observatory, P.O. Box 0, Soccoro, NM 87801, USA\\
$^{3}$Department of Physics and Electronics, Rhodes University, PO Box 94, Grahamstown, 6140, South Africa,\\
$^{4}$SKA South Africa, 3rd Floor, The Park, Park Road, Pinelands, 7405, South Africa
}

\date{Accepted XXX. Received YYY; in original form ZZZ}

\pubyear{2017}

\begin{document}
\label{firstpage}
\pagerange{\pageref{firstpage}--\pageref{lastpage}}
\maketitle

\begin{abstract}
We leverage the Sparsity Averaging Reweighted Analysis (SARA) approach for interferometric imaging, that is based on convex optimisation, for the super-resolution of Cyg A from observations at the frequencies 8.422GHz and 6.678{GHz} with the Karl G. Jansky Very Large Array (VLA). The associated average sparsity and positivity priors enable image reconstruction beyond instrumental resolution. An adaptive Preconditioned Primal-Dual algorithmic structure is developed for imaging in the presence of unknown noise levels and calibration errors. We demonstrate the superior performance of the algorithm with respect to the conventional \alg{clean}-based methods, reflected in super-resolved images with high fidelity. The high resolution features of the recovered images are validated by referring to maps of Cyg A at higher frequencies, more precisely 17.324GHz and 14.252GHz. We also confirm the recent discovery of a radio transient in Cyg A, revealed in the recovered images of the investigated data sets. Our \alg{matlab} code is available online on GitHub.
\end{abstract}


\section{Introduction}
New imaging techniques and algorithmic structures for radio interferometry have been extensively investigated in the recent years. The main objectives are: firstly to meet the next-generation instruments' capabilities in producing maps of the radio sky with unprecedented depth and resolution and secondly to cope with the sheer volume of the acquired data. Recently proposed compressive sensing techniques using convex optimisation for radio interferometric (RI) imaging have been shown to be very promising, potentially superseding the standard \alg{clean}-based techniques in terms of quality \citep[\emph{e.g.}][]{hogbom74,clark1980,csclean1983,Wakker88,Bhatnagar2004,Cornwell2008}, while, in principle, showing scalability to big data. The general approach consists in minimising a sum of convex functions. These include data fidelity terms and relevant regularisations for RI images, such as sparsity and positivity (typically for an intensity image). In particular, \cite{Carrillo2012} proposed the Sparsity Averaging Reweighted Analysis approach (SARA), where sparsity-by-analysis of the sky estimate is promoted in a collection of bases by solving consecutive re-weighted $\ell_1$ problems. Several algorithms based on convex optimisation have been proposed to solve the SARA minimisation problem, such as {Douglas-Rachford} splitting algorithm \citep{Carrillo2012} and the Simultaneous Direction Method of Multipliers (\ac{sdmm}) \citep{Carrillo2014}. \cite{onose2016} have proposed a Primal-Dual (\ac{PD}) algorithmic structure for RI imaging, in the context of which full splitting of the different functions involved is achieved, resulting in a highly parallelisable algorithm. 

More recently, \cite{onose2017} have proposed an accelerated Primal-Dual algorithmic structure reconciling two common data weighting schemes in RI imaging, namely natural and uniform weighting. On the one hand, natural weighting, by accounting only for the noise statistics, provides optimal sensitivity. On the other hand, uniform weighting, by additionally incorporating the density of the sampling, modifies the effective sampling and consequently, the associated point spread function (PSF), with the aim of optimising the resolution achieved within a finite number of iterations. Yet, this scheme reduces the overall sensitivity, since the scarcely sampled${-}$thus noisy${-}$measurements at the high spatial frequencies are over-weighted, while the highly sampled${-}$thus sensitive${-}$measurements at the low spatial frequencies are down-weighted. In the context of the \ac{PD} algorithm, the effect of the density of the sampling is instead cast in terms of the convergence speed.
The algorithmic structure proposed in \cite{onose2017}, dubbed Preconditioned Primal-Dual (\ac{ppd}) enforces data fidelity on the naturally-weighted data via non-Euclidean proximity operators, 
where projections onto the $\ell_2$ balls are generalised as projections onto $\ell_2$ ellipsoids incorporating the uniform weights. These ellipsoid projections provide accelerated convergence, thereby enabling simultaneous optimisation of the dynamic range and resolution of the recovered image. {In summary, the versatility of convex optimisation with respect to the choice of both the regularisation priors (\emph{e.g.} SARA) and the algorithmic structure (\emph{e.g.} \ac{PPD}) has opened the door to tremendous potential for enhancement of the quality of RI imaging when compared to \alg{clean}-based techniques.}

In this paper, we leverage the \ac{PPD} algorithmic structure for solving the SARA minimisation problem and provide high fidelity high resolution imaging of Cyg A from VLA observations at the frequencies $8.422\rm{GHz}$ and $6.678\rm{GHz}$. 
In addition to the thermal noise, the two data sets are corrupted with multiplicative calibration errors, which are likely due to the antennas' pointing errors. These are particularly important for strong radio sources such as Cyg A and indeed tend to dominate the thermal noise. We propose an \textit{adaptive} version of the \ac{PPD} algorithmic structure aiming for the estimation of the unknown levels of the noise and calibration errors in the data. On the one hand,  high fidelity maps of Cyg A at both frequencies are achieved thanks to the accurate estimation of the $\ell_2$ constraints on the data. On the other hand, super-resolved representations of Cyg A are obtained thanks to the average sparsity-promoting and positivity priors of SARA. Comparison with the standard Multi-Scale \alg{clean} algorithm (\ac{ms-clean}) \citep{cornwell08b} indicates the superior performance of the adaptive \ac{ppd} algorithm. Furthermore, the recent discovery of a faint transient radio source in the inner core of Cyg A, reported in \cite{perley2017}, is confirmed at the reconstructed maps of adaptive \ac{ppd}.

The remainder of the article is structured as follows. In Section \ref{sec:RI-CO}, we revisit the RI inverse imaging problem and the compressive sensing-based image reconstruction approach SARA. In addition, we briefly describe the \ac{PPD} algorithmic structure solving for SARA. In Section \ref{sec:PPD-RRI}, we present the adaptive \ac{ppd} algorithm. Cyg A imaging results from two data sets are presented in Section \ref{sec:SR-CYGA}. The achieved super-resolution with adaptive \ac{ppd} is analysed in Section \ref{ssec:sr}. We also report the detection of a secondary black hole in the super-resolved images of adaptive \ac{ppd}. Finally, conclusions are stated in Section \ref{sec:cend}.

~\\

\section{Sparse RI imaging to date}
\label{sec:RI-CO}
In this section, we revisit the RI measurement model and the minimisation problem to recover radio images, that is based on sparse representations and denoted \ac{SARA}. We also review the \ac{PD} algorithmic stucture solving the SARA minimisation problem, recently proposed in \cite{onose2017}.
 
\subsection{RI imaging problem}
RI data are Fourier measurements of the sky intensity modulated with the so-called Direction Dependent Effects (DDEs)\footnote{A particular case of DDEs are the so-called Direction Independent Effects (DIEs). These are constant complex-valued modulations in the image domain.}. These include the primary beam of the instrument and distortions induced by the propagation medium. { {Let $(\bs u, w)$ be the components of a baseline in units of the wavelength, with $w$ being the coordinate in the direction of the line of sight and ${\bs u}=(u, v)$ lying on its perpendicular plane. Assuming a monochromatic and non-polarised radiation, a RI measurement  $\ds V ({\bs u},w)$ reads }}
\begin{equation}
\label{eq:ri-contineous}
	\ds V(\bs{u},w) = \int G(\bs{l},{w}) I(\bs{l}) e^{-2 i \pi \bs{u} \cdot \bs{l}} \diff^2 \bs{l},
\end{equation}
where ${\bs l}$ are the coordinates of a radio point source in the plane tangent to the celestial sphere and  $\ds I({\bs l})$ is the unknown sky surface brightness at the position $\bs l$. $G(\bs{l},{w})$ stands for the DDEs, including the $w$-modulation that is resulting from the non-coplanarity of the radio interferometer and is given by $c(w,{\bs l})=e^{-2i\pi {w}(\sqrt{1- \vert {\bs l} \vert^2}-1)}$. The problem of recovering the sky intensity image from the radio measurements is an inverse problem and its discretised version reads
\begin{equation}
\label{eq:ri-discrete}
 \bs{y} = \bs{\Phi} \bs{x} + \bs{n} \rm{,~with~} \ds \bs{\Phi} = \bs{\Theta}\bm{G} \bm{F}{\bm Z}\bm{S} 
\end{equation}
where $\bs{x} \in \bb{R}_{+}^N$ is the intensity image of interest and $\bs{\Phi} \in \bb{C}^{M \times N}$ is the mapping operator from the image domain to the visibility space. The data $\bs y$ are the naturally-weighted RI measurements \emph{i.e.} $\bs y=\bs \Theta \bs{\tilde{y}}$, where $\bs {\tilde{y}}\in \bb{C}^M$ are the RI measurements and the operator
$\bs{\Theta} \in \bb{R}_{+}^{M \times M}$ accounts for the noise statistics and is a diagonal matrix whose elements are the square root of the natural weights. The operator $\bm{G} \in \bb{C}^{M \times o N}$ is the so-called gridding matrix, interpolating the RI measurements from the discrete Fourier components of the sky $\bs x$ which are lying on a regular grid.
Its rows are convolutional kernels, each centred at the corresponding spatial frequency $\bs u_{\ell\in \lbrace1,\dots,M\rbrace}$. $\bm{F} \in \bb{C}^{o N \times {o} N}$ is the Fourier matrix. $\bm Z$ is a zero-padding operator in the image space, allowing for a fine Fourier grid, thus a more accurate interpolation. $\bm S \in \bb{R}^{N \times N}$ is correcting for the convolution in the Fourier domain through $\bm G$. Note that, for the data investigated herein, the probed field-of-view is narrow (\emph{i.e.} $ \Vert  \bs{l}\Vert_2 \ll 1$). Thus, the $w$-modulation reduces to a flat function $ c(w,{\bs l})=1, \forall{\bs l}$. In general, it can be efficiently incorporated in the operator $\bm G$ as measurement-dependent convolutional kernels \citep{Dabbech17}.
\subsection{Sparse image reconstruction approach}
\label{ssec:sara}
Due to the incompleteness of the Fourier sampling and the presence of the noise, the problem of recovering the image of the sky $\bs x$ from the noisy measurements $\bs y$ is ill-posed. In order to reconstruct a reliable approximation, prior knowledge on the unknown sky is crucial and has to be considered in the imaging problem. In particular, the sparsity of the signal in adequate data representation spaces has been extensively adopted for RI imaging in the recent years \citep[\emph{e.g.}][]{Wiaux09,Li2011,Dabbech2012,Carrillo2012,Dabbech2015,Garsden2015}. Sparse regularisations are backed by the theory of compressive sensing \citep{Candes2006}. The theory proves that an exact recovery of the unknown signal can be achieved from noisy and incomplete measurements provided that the sensing basis $\bs{\Phi}$ is incoherent with the sparsity basis $\bs{\Psi}$ of the signal. Moreover, these regularisations can be easily enforced via convex functions. The resulting imaging problem can be efficiently solved using convex optimisation.

In the present work, we adopt the following minimisation problem, named the SARA approach, solving the inverse problem set in (\ref{eq:ri-discrete}) and originally proposed in \cite{Carrillo2012} 
\begin{eqnarray}
\label{eq:min-pb-0}
\min_{\bs x} \parallel {\bm W}{\bs \Psi}^\dagger {\bs x}\parallel_{1}
\rm{ s.t. } \left\{ \begin{aligned}
					\parallel{\bs y}-{\bs \Phi}{\bs x}\parallel_2 \leq \epsilon, \\
			{\bs x\geq \bm 0},
				  \end{aligned} \right.
\end{eqnarray} where $\epsilon$ is the $\ell_2$ norm of the noise and constitutes the bound on the data fidelity term. Sparsity of the unknown signal $\bs x$ is promoted by analysis, {\emph{i.e.}} its projection in a redundant data representation space is sparse. The adopted sparsity basis $\bs{\Psi}=\left[\bs \Psi_1,\dots, \bs\Psi_b\right] $ is a collection of nine orthogonal bases: the Dirac basis and the eight first Daubechies wavelet bases. The most intuitive measure of sparsity is the $\ell_0$ norm. However, being non-convex and yielding NP-hard problems, it is often replaced by its convex relaxation the $\ell_1$ norm. In SARA, a re-weighted $\ell_1$ norm is adopted, where the weighting matrix is $\bm{W}=[\bm{W}_{ 1}, \dots, \bm{W}_{b}]$, $\forall i\in\{1,\dots,b\}$, $\bm{W}_{i}\in \bb{R}_+^{N\times N}$ being diagonal matrices. Solving a sequence of re-weighted $\ell_1$ minimisation problems leads to nearly strictly-sparse signals \citep{Candes2008}. The SARA approach has been shown to provide superior imaging quality to \alg{clean}-based approaches both on simulations and few real data sets \citep{Carrillo2014,onose2016,pratley17A,onose2017}.

Data fidelity can be enforced in a distributed manner by splitting the data and the measurement operator into $d$ blocks as described in \cite{Carrillo2014, onose2016,onose2017}. In this setting, the minimisation task (\ref{eq:min-pb-0}) equivalently reads
\begin{eqnarray}
\label{eq:min-pb}
\min_{\bs x} \parallel {\bm W}{\bs \Psi}^\dagger {\bs x}\parallel_{1}
\rm{ s.t. } \left\{ \begin{aligned}
					 \parallel{\bs y}_j-{\bs \Phi}_j{\bs x}\parallel_2 \leq \epsilon_j, ~\forall j \in\{ 1,\dots ,d \},\\
			{\bs x\geq \bm 0},~~~~~~~~~~~~~~~~~~~~~~~~~~~~
				  \end{aligned} \right.
\end{eqnarray}
where for each $~j \in\{ 1,\dots ,d \}$, $\bs{\Phi}_j=\bs{\Theta}_j\bm{G}_j \bm{F}{\bm Z}\bm{S} $ is the measurement operator  associated with the data block $\bs{y}_j \in \bb{C}^{M_j}$. $\epsilon_j$ is the $\ell_2$ norm of the noise $\bs{n}_j \in \bb{C}^{M_j}$ and consequently the $\ell_2$ bound on the data block fidelity constraint. Note that the constraint formulation of the minimisation problem (\ref{eq:min-pb}) (and its equivalent formulation (\ref{eq:min-pb-0})) assumes accurate knowledge of the noise. This is challenging in real applications, in particular in the presence of significant calibration errors. These imply that the models of the operators ${\bs \Phi}_j$, $~j \in\{ 1,\dots ,d \}$, are approximate. The $\ell_2$ bounds $\lbrace\epsilon_j\rbrace_{~j \in\{ 1,\dots ,d \}}$ will therefore have to account not only for the thermal noise but also the calibration errors.

Different algorithmic structures based on convex optimisation have been adopted to address the minimisation problem (\ref{eq:min-pb}). These solvers fit within the proximal splitting methods \citep[see][for a review]{combettes09}. In this framework, a minimisation task is solved iteratively with each function handled individually. Typically, the differentiable functions are minimised using their gradient and the non-smooth functions are solved via their proximity operators.
\cite{Carrillo2014} adopted the Simultaneous Direction Method of Multipliers (\ac{SDMM}). The algorithm involves matrix inversions on the updates of the estimates of the solution. This results in a computational bottleneck when recovering large sized images, despite the separability of the different functions involved in the minimisation task. \cite{onose2016} proposed two algorithmic structures showing high scalability to big data. These are a sub-iterative variant of the Alternating Direction Method of Multipliers (\ac{admm}) and the \ac{PD} algorithm using forward-backward iterations. On the one hand, the \ac{admm}-based algorithmic structure presents a partial splitting of the functions involved. On the other hand, \ac{PD} allows for a full splitting of all the operators and functions. Furthermore, at each iteration, randomised updates on the different variables involved are allowed. The computational load per iteration is therefore reduced. The algorithm is also shipped with a preconditioning functionality bringing accelerated convergence \citep{onose2017}, hence the greater scalability of \ac{PD} to big data.
~\\

\subsection{The Primal-Dual algorithm}
In the PD algorithm, the following primal problem is solved
\begin{equation}
\label{eq:primal}
 \min_{\bs{x}} f(\bs{x}) + \gamma \sum_{i=1}^{b} l(\bm{W}_i\bs{\Psi}^\dagger_i\bs{x}) + \sum_{j=1}^{d} h_j(\bs{\Phi}_j\bs{x}),
\end{equation}
\\~
 together with its dual formulation
\begin{equation}
\label{eq:dual}
\min_{\substack{\bs{u}_i \\ \bs{v}_j}} f^* \Bigg(\!\!-\sum_{i=1}^b \bs{\Psi}_i \bm{W}_i \bs{u}_i - \sum_{j=1}^d \bs{\Phi}^\dagger_j \bs{v}_j \Bigg)+ \frac{1}{\gamma} \sum_{i=1}^b l^*(\bs{u}_i) + \sum_{j=1}^d h_j^*(\bs{v}_j).
\end{equation}
The parameter $\gamma$ is free and only affects the convergence speed, $\bs x $ is the primal variable that is the unknown image of the sky, and ${\bs{u}_{i \in \lbrace1,..,b\rbrace} ,~ \bs{v}_{j \in \lbrace1,..,d\rbrace}}$ are the dual variables associated with the sparsity priors and the data fidelity terms, respectively. The notation $(^*)$ stands for the Legendre-Fenchel conjugate function.
Note that, in the formulation (\ref{eq:primal}), further splitting of the sparsity prior with respect to each sparsity basis is achieved thanks to the separability of the $\ell_1$ norm. Moreover, the constraints are reformulated using the indicator function\footnote{Considering a convex set ${\mc{C}}$, its indicator function is defined as
\begin{equation}
	(\forall\bs{z}), ~\iota_{\mc{C}} (\bs{z}) \overset{\Delta}{=} \left\{ \begin{aligned}
					0 & \qquad \bs{z} \in \mc{C} \\
					+\infty & \qquad \bs{z} \notin \mc{C}
				 \end{aligned} \right. 
	\label{indicator-function}
\end{equation}}. The functions involved are $f = \iota_{{\mc{\bb{R}}^N_+}}$, enforcing the positivity and the reality of the unknown signal, 
$l = \| . \|_1$, imposing sparsity-by-analysis of the signal in the basis $\bs \Psi_i$, $\forall i\in\{1,\dots,b\}$ and 
$h_j = \iota_{\mc{B}_j} $, where ${\hquad \mc{B}}_j = \{ \bs{z} \in \bb{C}^{M_j}: \| {\bs{z} - \bs{y}}_j \|_2 \leq \epsilon_j \} $, are the data fidelity terms, enforcing the residual data blocks to be within the $\ell_2$ balls ${\mc{B}_j}$, $\forall j\in\{1,\dots,d\}$. The formulated problem (\ref{eq:primal}) is analogous to the problem (\ref{eq:min-pb}).

The different functions involved in (\ref{eq:primal}) and (\ref{eq:dual}) are non-differentiable, therefore they are minimised via their proximity operators. {Considering a lower semi-continuous and proper convex function $g$, its proximal operator is defined as 
\begin{equation}
	 (\forall\bs{z}), ~\prox_g (\bs{z}) \overset{\Delta}{=} \argmin_{\bar{\bs{z}}} g(\bar{\bs{z}}) + \frac{1}{2} \| \bs{z} - \bar{\bs{z}}\|_2^2.
	\end{equation}
Following this definition, the function $f$ is minimised via projections on the positive and real orthant, the sparsity function $l$ is minimised via soft-thresholding
operators and the data fidelity terms $h_j$ are minimised via projections
onto the $\ell_2$ balls ${\hquad \mc{B}}_j$, simultaneously. The proximal operators of the conjugate
functions involved in (\ref{eq:dual}) are obtained from those of (\ref{eq:primal}) by the
Moreau decomposition, as follows
\begin{equation}
	(\forall\bs{z}),~ \prox_{g^*} (\bs{z}) \overset{\Delta}{=} \bs{z} - \prox_g(\bs{z}).
	\end{equation}
The two minimisation tasks (\ref{eq:primal}) and (\ref{eq:dual}) are solved via forward-backward steps updating the dual and the primal variables \citep{Pesquet2015}. These consist in a gradient descent step coupled with a proximal update. In analogy with \alg{clean} \cite[\emph{e.g.} Cotton-Schwab \alg{clean}][]{csclean1983}, the algorithm can be understood as being composed of complex \alg{clean}-like steps performed in parallel in data, prior and image spaces \citep[the reader is directed to][for further details]{onose2016}.

\subsection{The Preconditioned Primal-Dual algorithmic structure}
\cite{onose2017} have recently proposed the algorithmic structure \ac{PPD}, where an acceleration strategy within the \ac{PD} algorithm is adopted. It consists of incorporating a priori knowledge on the data when enforcing fidelity to the naturally-weighted data. This is made feasible thanks to the generalised definition of the proximal operator which, considering a strongly positive self-adjoint linear operator $\bm U$, reads
\begin{equation}
(\forall\bs{z}),~\prox^{\bm{U}}_g (\bs{z}) \overset{\Delta}{=} \argmin_{\bar{\bs{z}}} g(\bar{\bs{z}}) + \frac{1}{2} (\bs{z} - \bar{\bs{z}})^\dagger \bm{U} (\bs{z} - \bar{\bs{z}}).
\end{equation}
Following this definition, for each data block indexed by $j \in\{1,\dots,d\}$, the $\ell_2$ projections are performed on skewed balls. Conceptually, these are equivalent to projections onto the ellipsoids ${\mc{E}_j}$, defined by $\mc{E}_j= \{ \bar{\bs{s}} \in \bb{C}^{M_j}: \| \bm{U}_j^{-\frac{1}{2}} \bar{\bs{s}} - \bs{y}_j \|_2 \leq \epsilon_j \}$, which are then moved to the $\ell_2$ balls ${\mc{B}_j}$ via the linear operator $\bm{U}_j^{-1/2}$ \citep[see][for further details]{onose2017}. In this setting, the operator $\bm U$ incorporates the prior information on the data and acts as a preconditioning matrix affecting only the speed of convergence while enforcing the data fidelity with respect to the naturally-weighted data. 
A relevant choice of the preconditioning matrix $\bm U$ to ensure a faster convergence involves the uniform weights \citep{onose2017}. More precisely, the matrix $\bm U\in \bb{R}_+^{M\times M}$ is set as a diagonal matrix, whose elements are inversely proportional to the density of the sampling at the vicinity of the probed Fourier modes. The more non-uniform the original density of the sampling over the Fourier plane, the more effective the approach.

The \ac{PPD} algorithmic structure is given in Algorithm \ref{alg-primal-dual}. Note that steps coloured in red are specific to adaptive \ac{ppd}, proposed in the following section. All the dual variables involved in the problem (\ref{eq:dual}) are updated in parallel via forward-backward steps. The dual variables associated with the data terms $\bs{v}_{j \in \lbrace1,..,d\rbrace}$ are updated in Step 5, where projections on the ellipsoids $\mc{E}_j$ are performed and the dual variables associated with the sparsity prior $\bs{u}_{i \in \lbrace1,..,b\rbrace}$ are updated via soft-thresholding operations controlled by the parameter $\kappa$ in Step 13. These dual variables are then utilised in Step 17 as incremental variables in the update of the primal variable $\bs x$, \emph{i.e.} the image of interest. The latter is followed with a projection onto the real positive orthant.

\begin{algorithm}[t]
\caption{{\bcs{Adaptive}} Preconditioned forward-backward \ac{pd}.}
\label{alg-primal-dual}
\begin{algorithmic}[1]
\small
\Given{$\bs{x}^{(0)}, \bar{\bs{x}}^{(0)}, \bs{u}_i^{(0)}, \bs{v}_j^{(0)}, \bm{W}_{ i}, \bm{U}_j, \epsilon_{j}^{{\bcs{(0)}}},\kappa, \tau, \eta, \zeta, {\bcs{\bs{\gamma},P, p_j^{(0)},\sigma^{(0)}}}$}

\RepeatFor{$t=1,\ldots$}
\Block{\bf run simultaneously}
\ParFor{$\forall j \in \lbrace 1,\dots,b\rbrace$}
	\State $\ds {\bs{v}}_j^{(t)} = {\bm{U}_j} \Bigg(\!\identity - {\bm{U}_j}^{-1/2} \proj_{\mc{E}_j^{{\bcs(t-1)}}}\!\!\! \Bigg) \! \Big(\bm{U}_j^{-1} \bs{v}_j^{(t-1)} \!+ \bs{\Phi}_j \bar{\bs{x}}^{(t)} \!\Big)$
  {\bcs{
  \State {${\mu}^{(t)}_j=\Vert\bs{y}_j-\bs{\Phi}_j {\bs{x}}^{(t)}\Vert_2$}

  \State{\bf{if}  $\sigma^{(t-1)}<\gamma_1$  \bf{and} $t-p^{(t-1)}_j\geq P $ \bf{and} $\frac{\vert \mu_j^{(t)} -\epsilon^{(t-1)}_j\vert }{ \epsilon^{(t-1)}_j} >\gamma_2$  }
 \State{~~~~~~$\epsilon^{(t)}_j = \gamma_3 \mu_j^{(t)}+(1-\gamma_3)\epsilon^{(t-1)}_j $}
 \State{~~~~~~$p^{(t)}_j=t$}
 \State{\bf{else set} $\epsilon^{(t)}_j=\epsilon^{(t-1)}_j$ and $p^{(t)}_j=p^{(t-1)}_j$ }
 }}

\EndParFor
	
\ParFor{$\forall i \in \lbrace 1,\dots,d\rbrace$}
	\State $\ds {\bs{u}}_i^{(t)} = \Bigg( \identity - \soft_{\kappa \|\bs{\Psi}\bm{W}\|_{\rm{S}}} \!\! \Bigg) \Big( \bs{u}_i^{(t-1)} + \bm{W}_{i}^\dagger \bs{\Psi}_i^\dagger \bar{\bs{x}}^{(t-1)} \Big)$
	
\EndParFor

\EndBlock{\bf end}
\State $\ds {\bs{x}}^{(t)} \! = \proj_{\mc{\mathbb{R}}_+^N} \! \Bigg(\! \bs{x}^{(t-1)} - \tau \Big(\! \eta  \! \sum_{j=1}^d \!   \bs{\Phi}^\dagger_j{\bs{v}}_j^{(t)} \! +\! \zeta \sum_{i=1}^b \! \bs{\Psi}_i \bm{W}_{i}{\bs{u}}_i^{(t)} \Big)~\!\!\Bigg)$
\State $\ds \bar{\bs{x}}^{(t)} = 2{\bs{x}}^{(t)} - \bs{x}^{(t-1)}$
\State{$\sigma^{(t)}=\frac{\Vert\ds {\bs{x}}^{(t-1)}-\ds {\bs{x}}^{(t)}\Vert_2}{\Vert\ds {\bs{x}}^{(t)}\Vert_2}$}
\Until {{\bf convergence}}
\Output{$\bs{x}^{(t)}, \bar{\bs{x}}^{(t)}, \bs{u}_i^{(t)}, \bs{v}_j^{(t)}$}
\end{algorithmic}
\end{algorithm}

\subsection{Re-weighted $\ell_1$ minimisation}
In order to achieve sparsity-by-analysis of the solution in the $\ell_0$ sense, consecutive re-weighted $\ell_1$ problems set in (\ref{eq:min-pb}) are solved in the Sparsity Averaging \emph{Re-weighted} Analysis approach (SARA) proposed in \cite{Carrillo2012,Carrillo2013}. We concisely re-explain the re-weighting procedure here for the sake of completeness. In this context, at each iteration indexed by $k$, a re-weighted $\ell_1$ minimisation problem associated with the weighting matrix $\bm W^{(k-1)}$ is solved using the \ac{PPD} algorithmic structure described in Algorithm \ref{alg-primal-dual}. The primal and dual variables involved in \ac{PPD} are initialised from the solutions of the previous weighted $\ell_1$ minimisation task. Once \ac{PPD} converges, the weighting matrix $\bm W^{(k)} = [\bm{W}^{(k)}_{ 1}, \dots, \bm{W}^{(k)}_{b}] $ is updated from the previous estimate of the primal variable $\bs{x}^{(k-1)}$ as follows
\begin{equation}
\op{D}_{e}\left(\bm{W}^{(k)}_{ i }\right) = \frac{\omega^{k}}{\omega^{k} +{ \alpha_i^{(k)}}{ \left( \left| \bm{\Psi}^\dagger_{ i} \bs{x}^{(k-1)} \right| \right)_{e}}},
\label{reweighting}
\end{equation}
with the operator $\op{D}_{ e}$ denoting the diagonal element $e$. The parameter $\omega$ is set such that $0<\omega <1$, ensuring the decrease of the weights and $\alpha_i^{(k)}$ is basis-dependent and is given by $\alpha_i^{(k)}=1/\max_e \left( \left| \bm{\Psi}^\dagger_{i} \bs{x}^{(k-1)} \right| \right)_{e}$, resulting in scale-free weights. Consequently, the weights are in the interval $[\frac{\omega^k}{2}~~1]$, that tends to $[0~1] $ asymptotically. Note that, for each $ i\in\{1,\dots,b\}$, the initial weighting matrix ${\bm W}_i^{(0)}$ is the identity matrix.
Given this definition, at each re-weighting step, the weights are decreased in such a way that significant analysis coefficients$-$corresponding to true signal$-$are strongly down-weighted. After several re-weights, their associated weights tend to zero. By doing so, only small-valued analysis coefficients$-$typically corresponding to noise$-$remain highly penalised by the $\ell_1$ norm (\emph{i.e.} their associated weights are close to 1). This weighting scheme is in line with the proposed scheme in \cite{Carrillo2012}. The iterative procedure, shown in Algorithm \ref{reweighted-pd}, stops when the relative variation between two consecutive estimates is within a bound $\varrho$ where $0<\varrho<1$ or the maximum number of iterations is reached.

\begin{algorithm}[t]
\caption{Re-weighting scheme.}
\label{reweighted-pd}
\begin{algorithmic}[1]
\small

\Given{$ \bs{x}^{(0)}, \bar{\bs{x}}^{(0)}, \bs{u}_i^{(0)}, \bs{v}_j^{(0)}, \bm{W}^{(0)}_{i}$}

\RepeatFor{$k=1,\ldots$}
\State {$
\left[\bs{x}^{(k)}, \bar{\bs{x}}^{(k)}, \bs{u}_i^{(k)}, \bs{v}_j^{(k)}\right ] = \mathrm{Algorithm~\ref{alg-primal-dual}} ~\big( \cdots \big) 
$}
\State {$\forall i \in \lbrace 1,\dots,d\rbrace$, {\bf update} $\bm{W}^{(k)}_i$}
\Until {{\bf convergence}}
\Output{$\bs{x}^{(k)}$}
\end{algorithmic}
\end{algorithm}

\section{Adaptive PPD for real imaging}
\label{sec:PPD-RRI}
Highly sensitive RI data from the new-generation arrays present prominent errors induced by the standard self-calibration, that is an iterative loop alternating between DIE calibration steps and \alg{clean} imaging steps. DIE modelling errors and lack of DDE calibration yield sky-dependent and correlated errors in the calibrated data. The effect of these errors is reflected in a 
limited dynamic range of the final recovered radio image. Joint DDE calibration and imaging as proposed in \cite{repetti17} would alleviate this effect drastically. However, this is out of the scope of the present article. Herein, the aim is imaging Cyg A from data calibrated with the standard RI pipelines. In this context, we assume that calibration errors share a common scale for each data snapshot, \emph{i.e.} data aggregated over a short time interval. The blocks in (\ref{eq:min-pb}) are therefore defined per snapshot and the associated $\ell_2$ bounds $\{\epsilon_j\}_{j\in\{1,\dots,d\}}$ will be set to account not only for the thermal noise but also the mismodelling of $\bs{\Phi}$ induced by calibration errors. The level of calibration errors per data block being a priori unknown, the bounds have to be estimated during image reconstruction. Note that when calibration errors are imperceptible (\emph{i.e.} buried in the thermal noise) the bounds on the data fidelity terms are fixed with respect to the statistics of the thermal noise \citep{Carrillo2012, onose2016}.

With the aim of posing the minimisation problem (\ref{eq:min-pb}) with the most appropriate $\ell_2$ constraints, we propose a strategy to adjust adaptively the $\ell_2$ bounds on the data fidelity terms during the iterations of the \ac{PPD} algorithm. The adaptive procedure described below is incorporated in \ac{PPD} through Steps 6-10 of Algorithm \ref{alg-primal-dual} (see modifications coloured in red). Technically, the original $\ell_2$ bounds $\{\epsilon_j\}_{j\in\{1,\dots,d\}}$ become iteration-dependent $\{\epsilon^{(t)}_j\}_{j\in\{1,\dots,d\}}$, $t$ being the iteration's index. At each iteration, $\epsilon_j^{(t)}$ is updated as a weighted mean of the current $\ell_2$ bound $\epsilon_j^{(t-1)}$ and the $\ell_2$ norm of the associated residual data ${\mu}^{(t)}_j=\Vert\bs{y}_j-\bs \Phi_j {\bs{x}}^{(t-1)}\Vert_2$ (Step 8 in Algorithm \ref{alg-primal-dual}). These updates are performed when the following conditions are met. (i) The estimate of the sky saturates, \emph{i.e.} the relative variation between two consecutive estimates $\sigma^{(t-1)}$ (calculated in Step 18) is below a fixed value $\gamma_1$. (ii) A minimum number of iterations is performed between two consecutive updates of the $\ell_2$ bound. (iii) 
The relative difference between the current estimate of the $\ell_2$ bound $\epsilon^{(t-1)}_j$ and the $\ell_2$ norm of the corresponding residual data ${\mu}^{(t)}_j$ is above a certain bound $\gamma_2$, where $0<\gamma_2<1$. If the the data block does not satisfy its $\ell_2$ constraint defined by $\epsilon^{(t-1)}_j$, the latter is assumed under-estimated and is increased. Otherwise, it is considered over-estimated and is therefore decreased. These conditions are checked at each iteration independently for all the data blocks. Conceptually, the update of the $\ell_2$ constraints redefines the minimisation problem posed in (\ref{eq:min-pb}). In this context, conditions (i) and (ii) are set to avoid the early modification of the posed minimisation problem, thus ensuring the stability of the strategy. 

To initialise the bounds, we perform imaging with the Non-Negative Least Squares algorithm (NNLS). For each data block indexed by ${j\in \{1,\dots,d \} }$, we first compute $\tilde{\bs{x}}_j^{\rm{NNLS}}$, that is given by
\begin{equation}
\tilde{\bs{x}}_j^{\rm{NNLS}}=\argmin_{\bs x} \parallel{\bs y}_j-{\bs \Phi}_j{\bs x}\parallel_2^2, ~\rm{s.t.~} {\bs x\geq \bm 0}.
\end{equation}
We then set the $\ell_2$ bound $\epsilon_j^{(0)}$ to $\mu_j^{\rm{NNLS}}=\| \bs{y}_j-\bs{\Phi}_j\tilde{\bs{x}}_j^{\rm{NNLS}}\|_2$. 
 Since only positivity is imposed, the NNLS minimisation problem is under-regularised and the model image tends to over-fit the noisy data. As a consequence, the bounds $\{\epsilon_j^{(0)} \}_{j\in \{1,\dots,d \} }$ tend to be highly under-estimated. Given this initialisation, in adaptive \ac{PPD}, the bounds are adaptively increased while enforcing sparsity. The saturation of the estimate of the solution and consequently the estimates of the bounds are highly correlated with the soft-thresholding parameter $\kappa$, inducing sparsity (see Step 13 of Algorithm \ref{alg-primal-dual}). In fact, when $\kappa$ is chosen too small, the estimate of the solution converges rapidly in $\ell_2$ balls whose bounds are very close to $\{\mu_j^{\rm{NNLS}}\}_{j\in\{1,\dots,d\}}$. In this case, the solution is under-regularised and noisy. Whereas, when $\kappa$ is chosen too high, the estimate of the solution converges in $\ell_2$ balls whose bounds are significantly higher than the noise level. In this case, the solution is over-regularised and too sparse. In \cite{onose2016}, the scale-free parameter $\kappa$ is advised to be set within the interval limited by $10^{-5}$ and $10^{-3}$, that is also in line with \cite{Carrillo2014}. Though this range remains relevant for adaptive \ac{ppd}, the algorithm is more sensitive to the choice of the soft-thresholding parameter due to the estimation of the $\ell_2$ bounds. For the data imaged herein, a value of order $10^{-5}$ is found to yield good results.

As detailed above, the resulting algorithmic structure, dubbed adaptive \ac{PPD}, is very similar to \ac{PPD} except for its additional feature, which is the $\ell_2$ bounds adjustments. An overview of the variables and parameters associated with the adaptive procedure is provided in Appendix \ref{apx:a}. 
Formally, \ac{PPD} is solving the minimisation problem set in (\ref{eq:min-pb}) with well-defined $\ell_2$ constraints. While, adaptive \ac{ppd} is solving consecutive minimisation tasks each corresponding to a different set of $\ell_2$ bounds. In fact, if the conditions (i), (ii) and (iii) are met for at least one data block, the adjustment of its associated $\ell_2$ bound is performed, hence a new minimisation problem is posed and solved, with all variables involved initialised from the last estimates of the previous minimisation task. The adaptive \ac{ppd} algorithm converges when the estimate of the sky saturates and all the $\ell_2$ constraints are satisfied. In this case, the weighting matrix involved in the sparsity prior is updated and a new iteration in Algorithm \ref{reweighted-pd} is performed, where a re-weighted $\ell_1$ minimisation task is solved with adaptive \ac{PPD} given the current estimates of the $\ell_2$ bounds.

\section{Cyg A Imaging with adaptive \ac{ppd}}
\label{sec:SR-CYGA}
VLA data sets investigated herein consist of aggregated data acquired with the four configurations of the instrument. The resulting Fourier sampling of the combined data is highly non-uniform. In this case, the preconditioning strategy in the adaptive \ac{PPD} algorithm is highly effective. Furthermore, since the observations were taken on four different days over a span of over one year, the noise statistics and calibration errors are not consistent over the whole data set. Consequently, assigning different bounds on the data blocks in comparison with assigning one global bound on all the data is crucial. In this section, we present the maps of Cyg A at two frequencies imaged with adaptive \ac{ppd}. We show the efficiency of the proposed algorithmic structure in recovering superior representations of the radio sky in comparison with the conventional approach \ac{ms-clean} \citep{cornwell08b}. 

\subsection{VLA observations}
The data under scrutiny are part of wide-band VLA observations of the well-studied radio galaxy Cyg A within the frequency range 2-18~GHz, performed over two years (2015-2016). {{The data sets correspond to observations at{  {X band ($8-12$ GHz)}} centred at the frequency $8.422\rm{GHz}$ and { {C band ($4-8$ GHz)}} centred at the frequency $6.678\rm{GHz}$, each over a spectral window of 128MHz and with a spectral resolution of 2MHz}}. The phase center is given by $\rm{RA}=19\rm{h}~59\rm{mn}~28.356\rm{s}$ ($J2000$) and $\rm{DEC}=+40^{\circ}~44\arcmin~2.07\arcsec$. All four configurations (A, B, C and D) of the VLA have been employed. Their respective total integration times associated with the two data sets are displayed in Table \ref{tab:times}.  {{In both observations, the initial time averaging is of 2 seconds. Decimation of the data sizes is performed via time and frequency averaging over 10 seconds and 8MHz. The data sets processed herein are of sizes $2\times 10^6$ and $1.3\times 10^6$ for X band and C band, respectively. }} 
They have been carefully calibrated using well-established techniques in \alg{aips}, consisting of iterative self-calibration alternating between DIE calibration steps and Cotton-Schwab \alg{clean} imaging steps \citep{csclean1983}. No DDE calibration is performed, hence the dynamic range on the recovered maps is constrained by the subsequent artefacts rather than the thermal noise.

 	\begin{table}
 	
	\centering
	\caption{Total integration times for the data sets at the two frequencies.}
	\label{tab:times}
	\begin{tabular}{lcccr} 
		Array configuration & A & B& C & D\\
		\hline
		8.42GHz~ & $11.38\rm{h}$&$3\rm{h}$& $1.63\rm{h}$ & $0.80\rm{h}$\\
		6.67GHz & $6.11\rm{h}$ &$2.01\rm{h}$  & $1.63\rm{h}$ &$0.58\rm{h}$       \\
		
		\hline
	\end{tabular}
\end{table}

\subsection{Imaging quality assessment}
\label{ssec:imaging}
To assess the quality of the reconstructions, we perform visual inspections of the obtained images. These are the estimated model image $\tilde{\bs x}$ and the residual image $\bs r = \beta \bs{\Phi}^\dagger(\bs y-\bs{\Phi}\tilde{\bs x})$, where $\beta$ is a normalisation factor\footnote{Here, we adopt the conventional normalisation of the residual image in RI. That is scaling the residual $\bs{\Phi}^\dagger(\bs y-\bs{\Phi}\tilde{\bs x})$ by $\beta=1/\max_i(\bs{\Phi}^\dagger\bs{\Phi}{\bs{\delta}})_{i} $, where $\bs\delta$ is an image with value 1 at the phase center and zero otherwise. By doing so, the PSF defined as ${\bs{h}}=\beta\bs{\Phi}^\dagger\bs{\Phi}{\bs{\delta}}$ has a peak value equal to 1.}.
In the context of imaging with the \alg{clean}-based technique \ac{MS-clean}, we also consider the restored image ${\bs z}= \tilde{\bs x}\ast {\bs b}+{\bs r}$, that is the estimated model image convolved with the so-called \alg{clean} beam ${\bs b}$, typically a Gaussian fitted to the PSF's primary lobe, and to which the residual image $\bs r$ is added. Convolving the estimated model image with the \alg{clean} beam is standard in \alg{clean} imaging. The latter assumes that the sky is made of point sources. Hence, the obtained model image consists of gridded point sources, many of which can be$-$indeed, need to be$-$negative. This translates in over-emphasising of the high spatial frequency content of the recovered model image.
Such a model of the radio sky is physically unreasonable. Therefore, the standard recourse is to smooth the image.
By doing so, a more physical representation of the radio image at the resolution of the instrument is obtained.
This is not required for compressive sensing-based approaches. Thanks to the use of both more complex and physical regularisations and explicit data fidelity bounds (see (\ref{eq:min-pb})), these approaches have been shown to achieve accurate estimates of the ground truth images in synthetic observations and good approximations of the true sky in early real applications \citep{Wiaux09b,wenger10,Carrillo2012,Dabbech2015, Garsden2015,onose2016,pratley17A}. {Moreover, recent studies have shown that applying a restoring beam on the reconstructed images with this class of methods does not enhance the fidelity to the ground truth image \citep{Chael16,Akiyama17} as opposed to \alg{clean}-based methods.} Hence, no post-processing convolution by a \alg{clean} beam or addition of the residual image is recommended.  The RI recovered image in the context of compressive sensing-based approaches is the estimated model image.

To quantify the performance of the imaging techniques, we consider the dynamic range metric, which is often adopted in RI imaging and is defined as $DR=\max_i{{ z}_i}/\sigma_{\bs r}$, where $\sigma_{\bs r}$ is the standard deviation of the residual image ${\bs r}$ and $\bs z$ is the restored image as defined in the context of \alg{clean} imaging. In the computation of the $DR$ values for adaptive \ac{ppd}, the involved images are obtained as follows. For natural weighting, we compute the residual image $\bs{r}^{\rm{\ac{ppd}}} = \beta \bs{\Phi}^\dagger(\bs y-\bs{\Phi}\tilde{\bs x}^{\rm{\ac{ppd}}})$ 
and the image ${\bs z}^{\rm{\ac{ppd}}}= \tilde{\bs x}^{\rm{\ac{ppd}}}\ast {\bs b}+{\bs r}^{\rm{\ac{ppd}}}$. 
For Briggs weighting, let $\bar{\bs y}=\bar{\bs{\Theta}}\tilde{\bs{y}}$ and $\bar{\bs{\Phi}}=\bar{\bs{\Theta}}\bm{G} \bm{F}{\bm Z}\bm{S}$ denote the Briggs-weighted data and their associated measurement operator, where $\bar{\bs{\Theta}}$ is a diagonal matrix whose elements are the square root of the Briggs weights. We compute the residual image $\bar{\bs{r}}^{\rm{\ac{ppd}}} = \bar{\beta} \bar{\bs{\Phi}}^\dagger (\bar{\bs y}-\bar{\bs{\Phi}}\tilde{\bs x}^{\rm{\ac{ppd}}})$ 
and the image $\bar{\bs z}^{\rm{\ac{ppd}}}= \tilde{\bs x}^{\rm{\ac{ppd}}}\ast \bar{\bs b}+{\bar{\bs r}}^{\rm{\ac{ppd}}}$. The kernels ${\bs b}$ and ${\bar{\bs b}}$ are the \alg{clean} beams associated with natural and Briggs weighting schemes, respectively. 
However, the $DR$ metric may not reflect accurately the dynamic range in the restored image, since by definition, it is biased by the residual image. In fact, a residual image with a low standard deviation can be induced by false detections in the model image, in particular when positivity is not imposed. This is often the case of \alg{clean}-based algorithms. Therefore, we report an alternative definition of the dynamic range based on the model image solely, that we call model dynamic range $MDR=\max_i{\tilde{ x}_i}/{\tilde{ x}_k}$, where ${\tilde{ x}_k}$ is the brightest pixel corresponding to an artefact in the estimated model image\footnote{ {The pixel position is determined through the visual inspection of the model image as the one with the highest pixel value and not belonging to the support of the source.}}. Such a metric is relevant for compressive sensing approaches where the estimated model images are characterised with realistic surface brightness, in particular positive. We do not report $MDR$ for the \ac{ms-clean} model maps as these exhibit unrealistic features, in particular prominent negative components. 

Due to the absence of the ground truth image of the sky, we examine the smoothed versions of the estimated model images at the resolution of the instrument. To compare adaptive \ac{ppd} with naturally-weighted \ac{ms-clean}, we adopt the images $\tilde{\bs z}=\tilde{\bs x}\ast {\bs b}$. Similarly, we adopt the images $\bar{\tilde{\bs z}}=\tilde{\bs x}\ast \bar{\bs b}$ to compare adaptive \ac{ppd} with Briggs-weighted \ac{ms-clean}. We assess the similarity of these two sets of images using the following metric, defined for two signals ${\bs x}_{1}$ and ${\bs x}_{2}$ as $S({\bs x}_1,{\bs x}_2)=20\log_{10}( \max(\parallel {\bs x}_1\parallel_2,\parallel {\bs x}_2\parallel_2)/\parallel {\bs x}_1-{\bs x}_2\parallel_2)$. We re-emphasise that smoothing the model image obtained with adaptive \ac{ppd} is not recommended and is performed here only for comparison purposes with \ac{ms-clean}.
\subsection{Imaging results}
\label{ssec:rslts}
The performance of adaptive \ac{ppd} is evaluated in comparison with the standard RI imaging technique \ac{ms-clean} with two weighting schemes: natural and Briggs. The latter weighting scheme constitutes a compromise between uniform and natural weighting, controlled by a robustness parameter. It is chosen herein over uniform weighting, in the aim to present the optimal reconstructions with \ac{Ms-clean}. On a further note, adaptive \ac{ppd} imaging is performed in \alg{matlab} and \ac{MS-clean} imaging is performed using the RI imaging software \alg{wsclean} \citep{Offringa2014}.
\subsubsection*{X band}
\label{ssec:xresults}
The imaged radio map from the data at a frequency of 8.422GHz is of size $4096\times2048$ pixels, with a pixel size $\delta l=0.04\arcsec$ (in both directions). The chosen pixel size is such that $\delta l =1/5B_X$, where $B_X=\max\limits_{\bs u_{\ell\in \lbrace1,\dots,M\rbrace}}\Vert \bs{u}_{\ell}\Vert_2$ is the maximum baseline \emph{i.e.} the spatial band-limit of the observations. This corresponds to recovering the signal up to 2.5 times the nominal resolution \emph{i.e.} its recovered spatial bandwidth is $\tilde{B}_X\approx2.5\times B_X$. Such a choice of the imaging resolution is conventional in RI imaging\footnote{In order to have reasonable results with \alg{clean}, the PSF needs to be adequately sampled. Therefore, it is common in RI imaging to set the pixel size $\delta l$ such that $1/5B_X\leq\delta l \leq1/3B_X$.}. We split the data to 22 blocks of size $10^5$ measurements on average, {{where each block is a single snapshot \emph{i.e.} data acquired within a time interval over which certain errors (like pointing offsets), can be assumed constant}}. The number of blocks is chosen to take advantage of the parallelised structure of adaptive \ac{PPD}. We perform 70 weighted $\ell_1$ minimisation tasks using adaptive \ac{ppd}. Each minimisation task stops when the relative variation between two successive estimates of the sky is below $10^{-5}$. To ensure higher accuracy of the final solution, the last minimisation task stops when the relative variation between two successive estimates of the sky is below $10^{-6}$. For \ac{ms-clean}, we consider imaging with the weighting schemes: natural and Briggs (the robustness parameter is set to $r=-1$). 
We re-emphasise that in imaging with adaptive \ac{ppd}, only natural weighting$-$consisting in whitening the noise$-$is applied on the measurements in order to reach the optimal sensitivity. 

\begin{figure*}
\centering
\begin{minipage}[t]{1\linewidth}
\vspace{0pt} 
\centering
\includegraphics[width=0.355\linewidth]{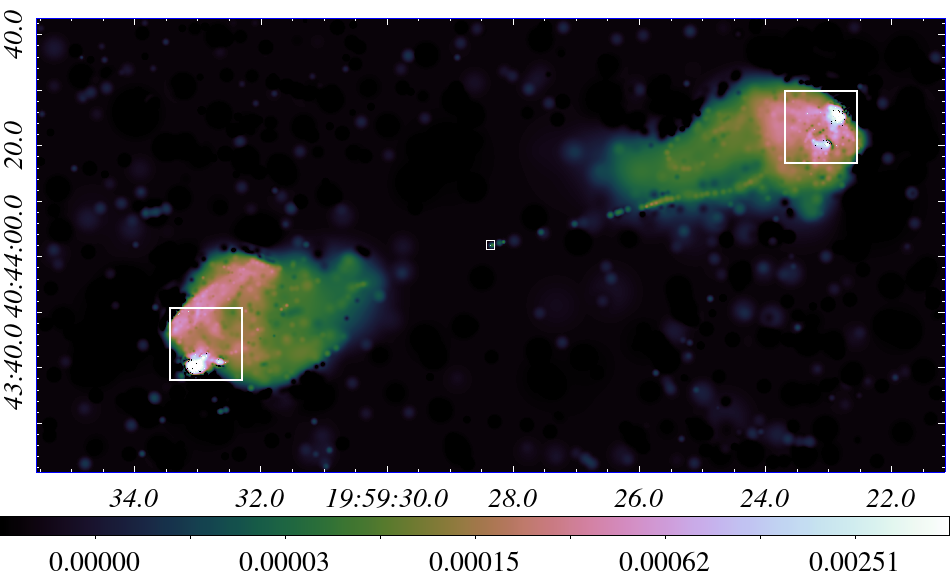}
\vspace{0pt} 
\includegraphics[width=0.201\linewidth]{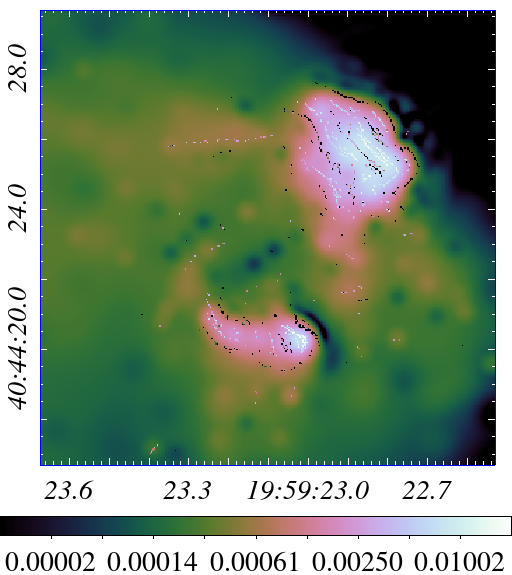}
\vspace{0pt} 
\includegraphics[width=0.201\linewidth]{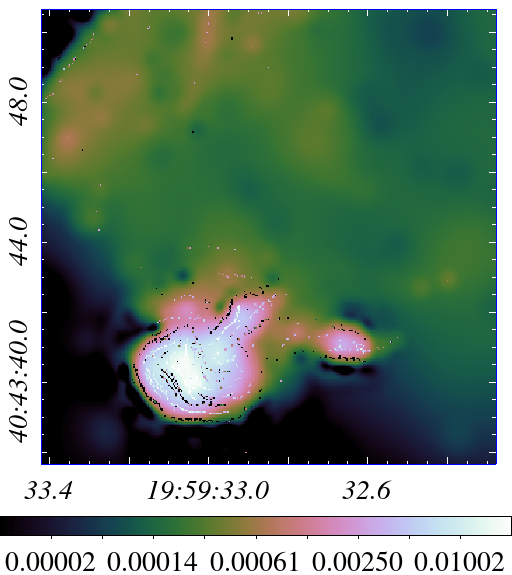}
\vspace{0pt} 
\includegraphics[width=0.201\linewidth]{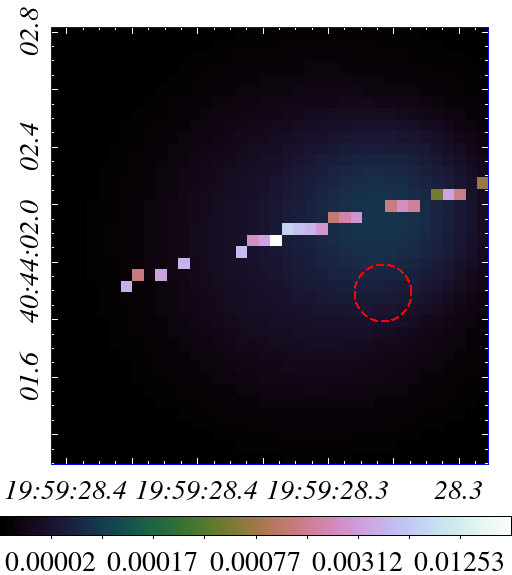}
\end{minipage}
\begin{minipage}[t]{1\linewidth}
\vspace{0pt} 
\centering
\includegraphics[width=0.355\linewidth]{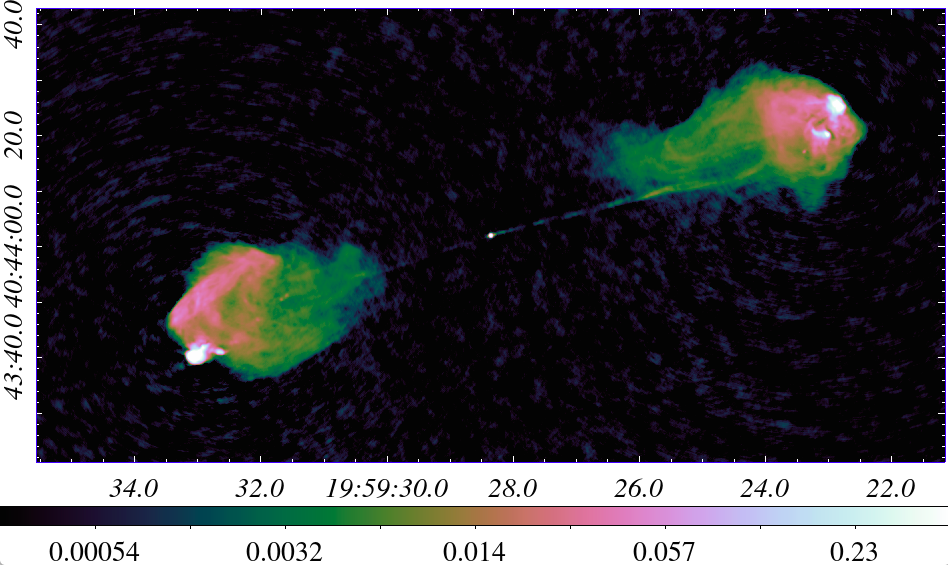}
\vspace{0pt} 
\includegraphics[width=0.201\linewidth]{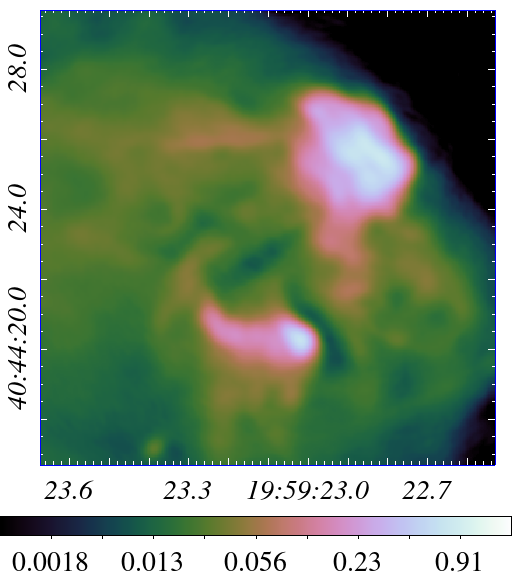}
\vspace{0pt} 
\includegraphics[width=0.201\linewidth]{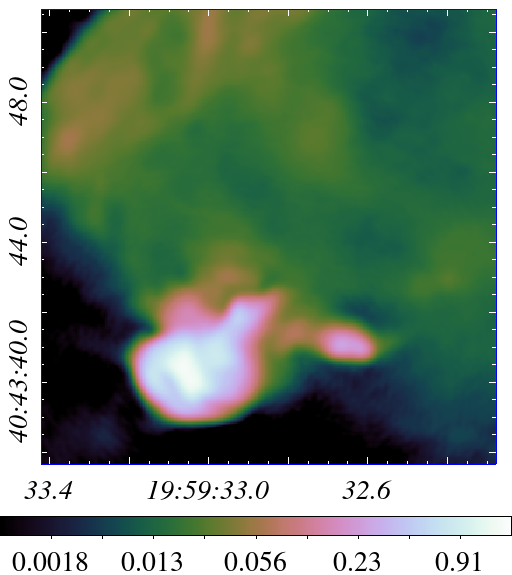}
\vspace{0pt} 
\includegraphics[width=0.201\linewidth]{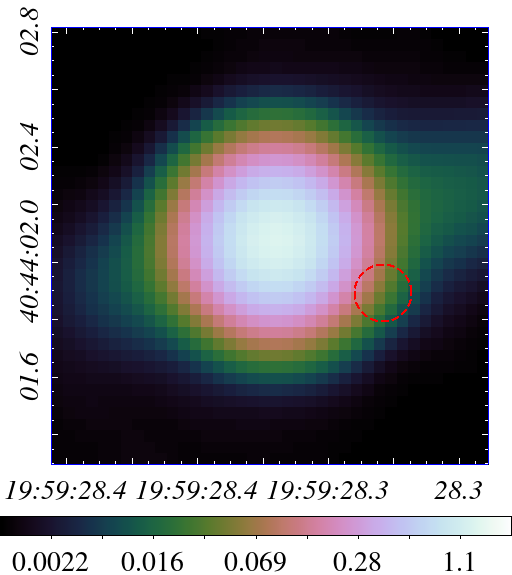}
\end{minipage}
\begin{minipage}[t]{1\linewidth}
\vspace{0pt} 
\centering
\includegraphics[width=0.355\linewidth]{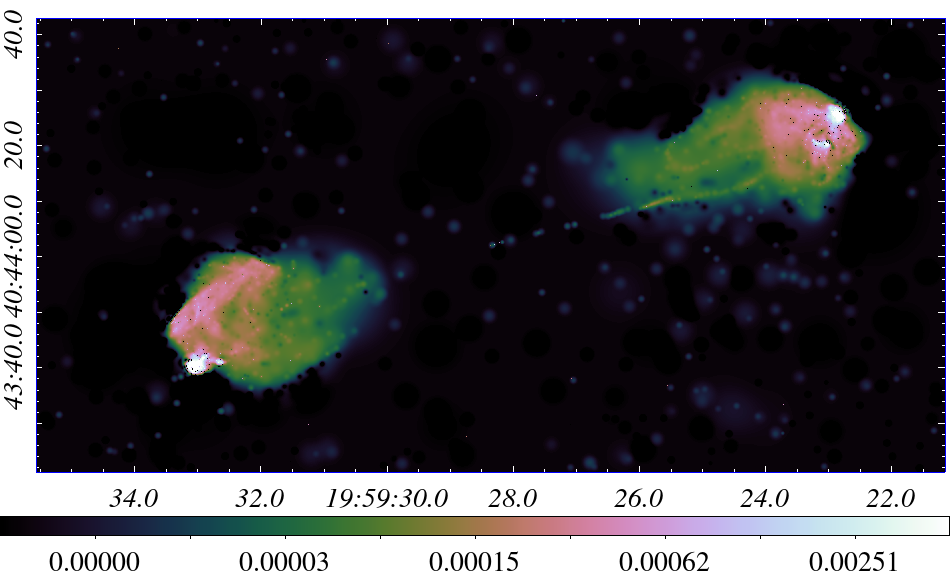}
\vspace{0pt} 
\includegraphics[width=0.201\linewidth]{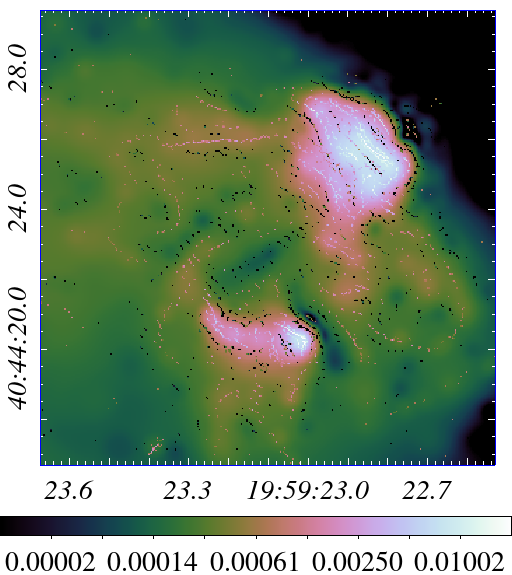}
\vspace{0pt} 
\includegraphics[width=0.201\linewidth]{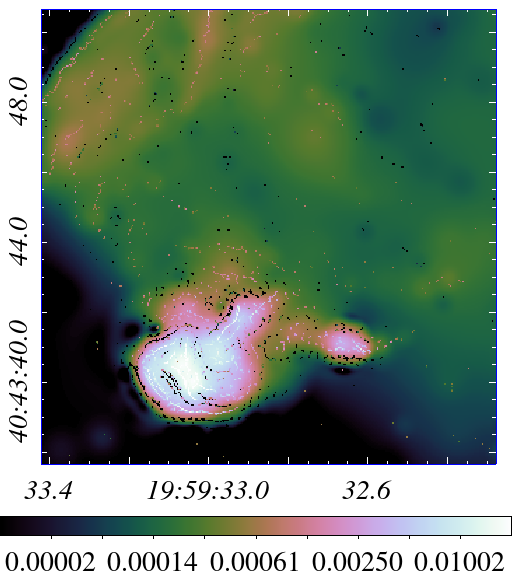}
\vspace{0pt} 
\includegraphics[width=0.201\linewidth]{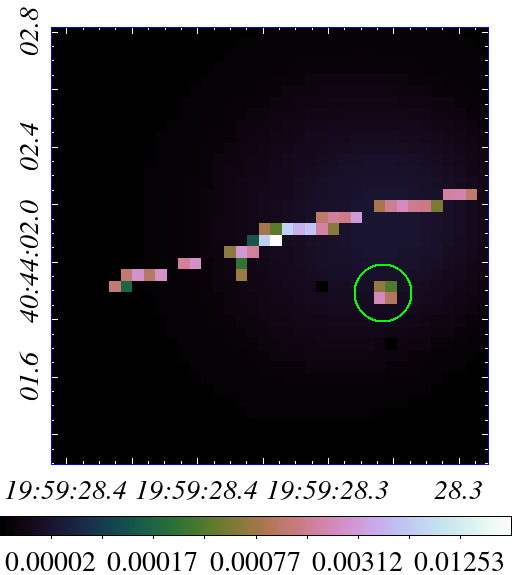}
\end{minipage}
\begin{minipage}[t]{1\linewidth}
\vspace{0pt} 
\centering
\includegraphics[width=0.355\linewidth]{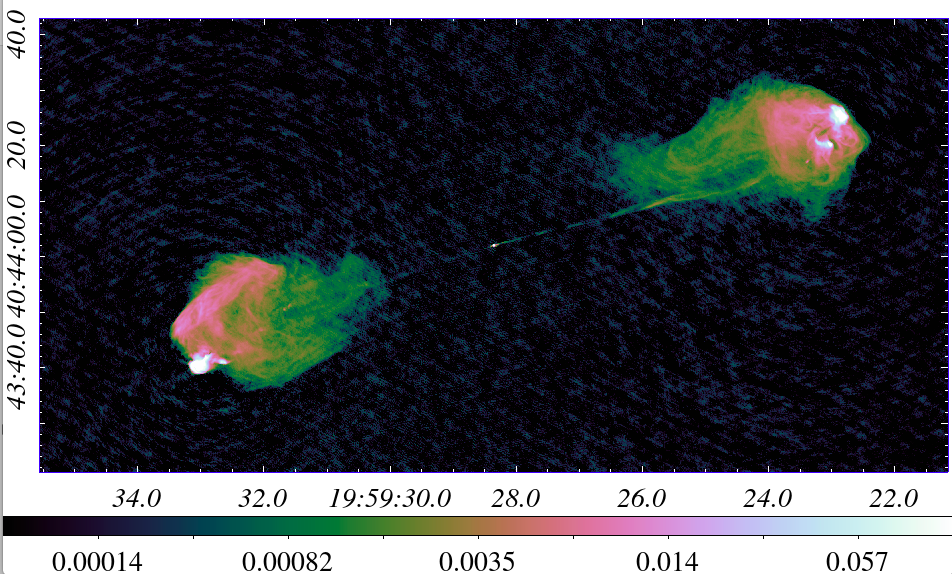}
\vspace{0pt} 
\includegraphics[width=0.201\linewidth]{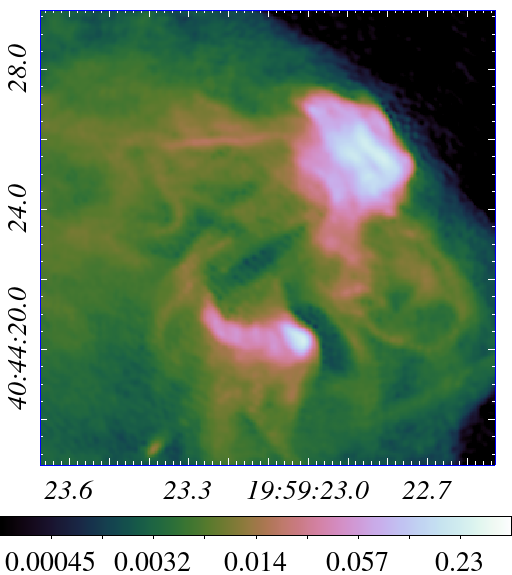}
\vspace{0pt} 
\includegraphics[width=0.201\linewidth]{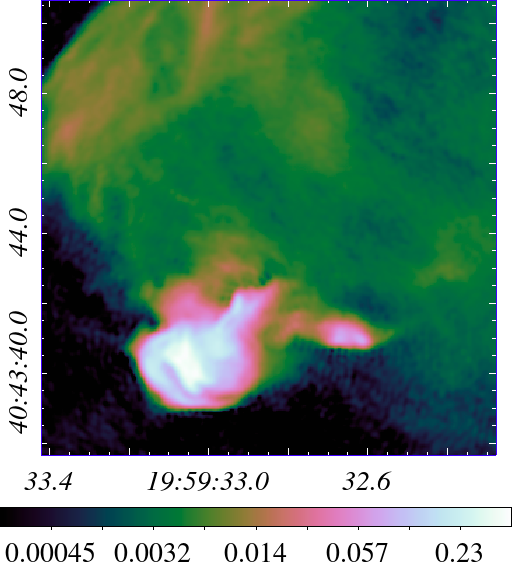}
\vspace{0pt} 
\includegraphics[width=0.201\linewidth]{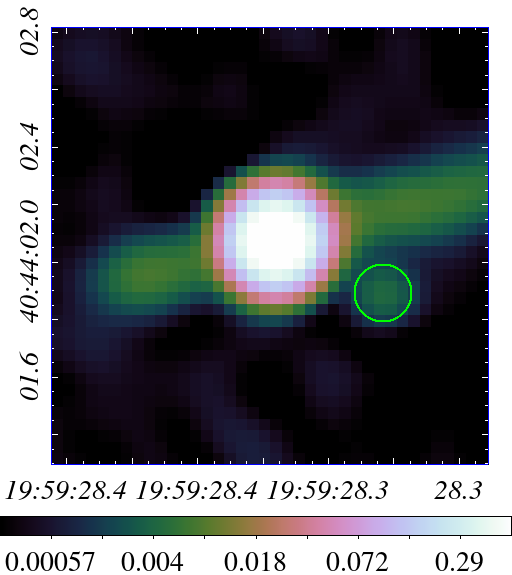}
\end{minipage}
\begin{minipage}[t]{1\linewidth}
\vspace{0pt} 
\centering
\includegraphics[width=0.355\linewidth]{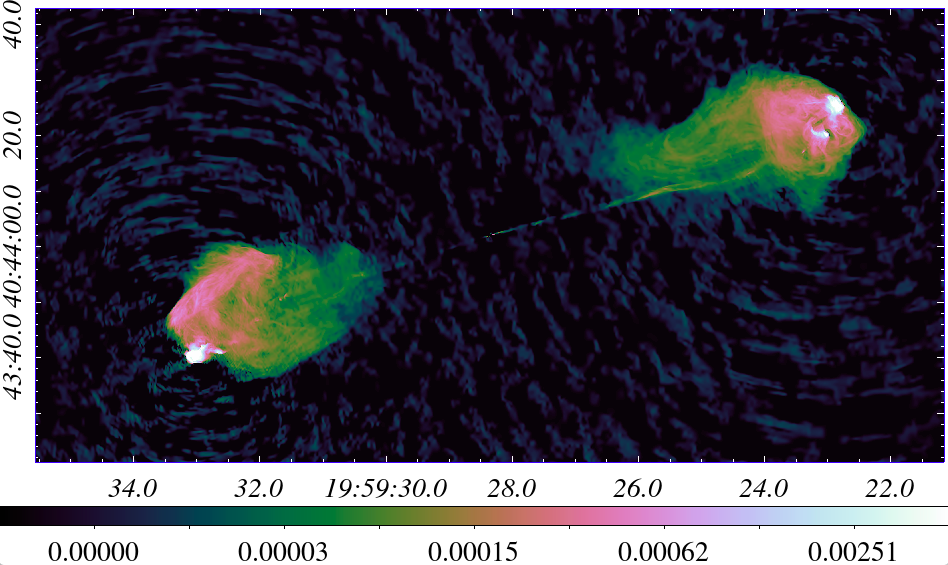}
\vspace{0pt} 
\includegraphics[width=0.201\linewidth]{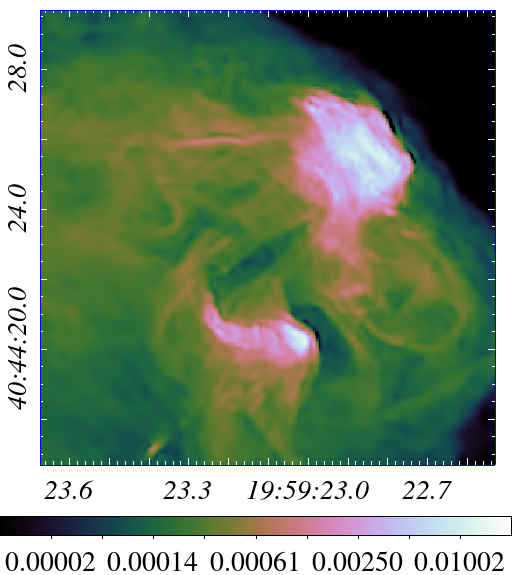}
\vspace{0pt} 
\includegraphics[width=0.201\linewidth]{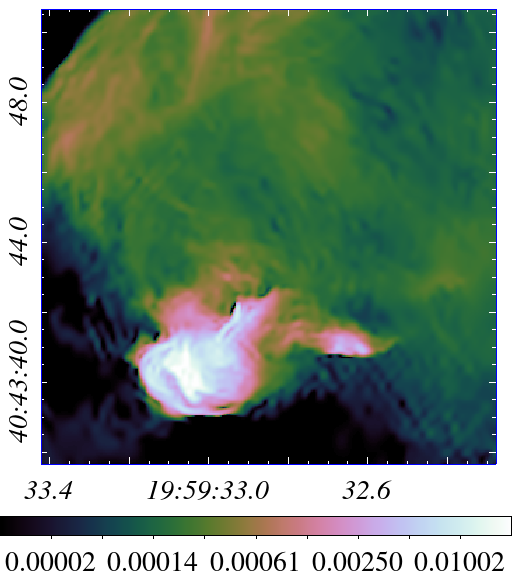}
\vspace{0pt} 
\includegraphics[width=0.201\linewidth]{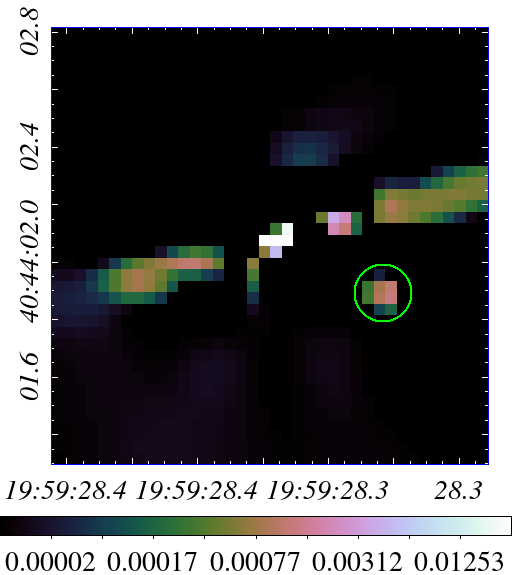}
\end{minipage}
\begin{minipage}{1\linewidth}

\caption{
\label{fig:xmzooms}
X band: recovered images at $2.5$ times the resolution of the instrument. { {From top to bottom, estimated model and restored images of naturally-weighted \ac{ms-clean} (resp. rows 1 and 2), estimated model and restored images of Briggs-weighted \ac{ms-clean} (resp. rows 3 and 4) and model image of adaptive \ac{ppd} (fifth row). The full images are displayed in $\log_{10}$ scale (first column) as well as zooms on the three brightest regions: east jet's hotspot (second column), west jet's hotspot (third column) and the inner core of the Cyg A galaxy (fourth column). The zoomed regions are highlighted with white boxes in the model image of naturally-weighted \ac{ms-clean} (top row, left column). The surface brightness of the restored image obtained with naturally-weighted \ac{ms-clean} (second row) is in units of ${\rm{Jy}}/{\rm{beam}}$, the naturally-weighted beam is of size $0.35\arcsec \times 0.35\arcsec$ and its flux is $90.43{\rm{Jy}}$. The surface brightness of the restored image obtained with Briggs-weighted \ac{ms-clean} (fourth row) is also in units of ${\rm{Jy}}/{\rm{beam}}$, the Briggs-weighted beam is of size $0.18\arcsec \times 0.18\arcsec$ and its flux is $22.95{\rm{Jy}}$. The surface brightness of the model images (rows 1, 3 and 5) is in units of ${\rm{Jy}}/{\rm{pixel}}$, the pixel size being $0.04\arcsec$ in both directions.}} Note that the black dots in the hotspots recovered in the model images of \ac{ms-clean} correspond to important negative pixels. Naturally, these are non-physical components for an intensity map. Therefore, astronomers utilise instead the restored maps, where the prominent negative components disappear thanks to the blurring of the model image with the \alg{clean} beam.}
\end{minipage}
\end{figure*}

\begin{figure*}
\vspace{0pt}
\includegraphics[width=0.3\linewidth]{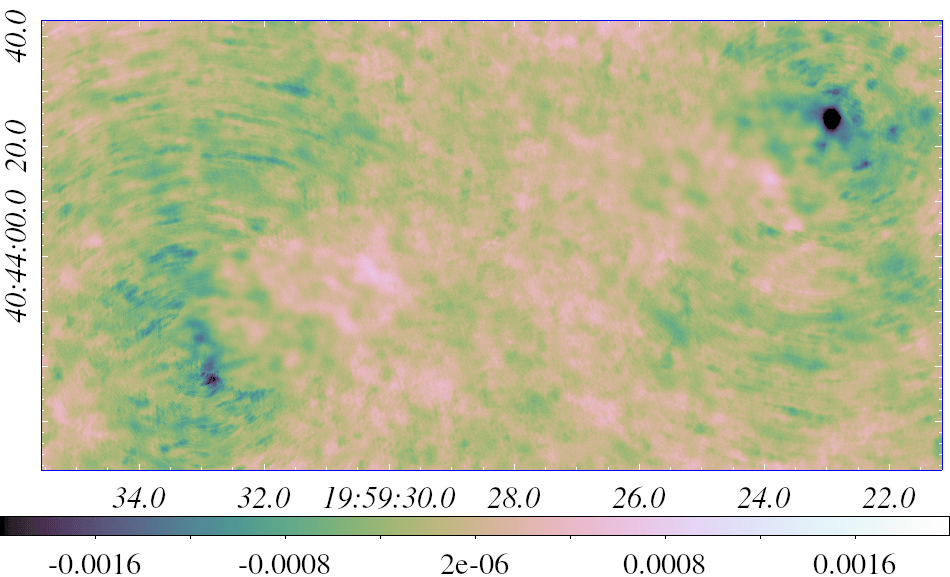}
\vspace{0.0pt} 
\includegraphics[width=0.3\linewidth]{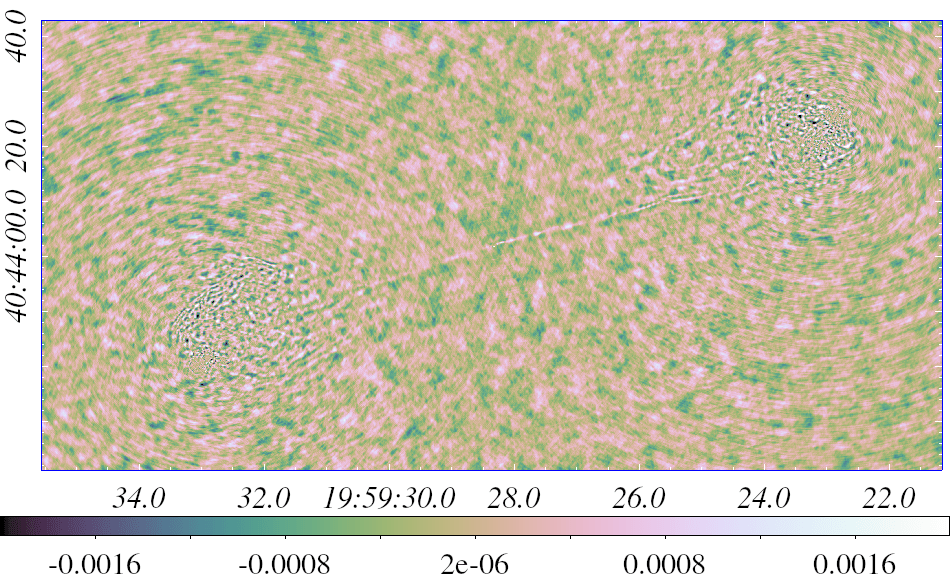}
\vspace{0.0pt} 
\includegraphics[width=0.3\linewidth]{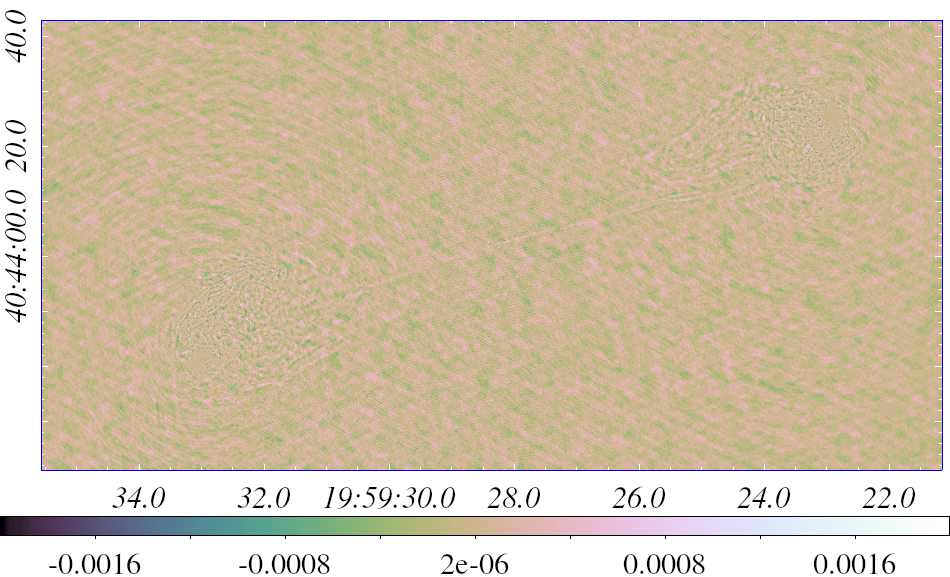}
\caption{
\label{fig:xres} 
{X band: residual images. From left to right: residual image of adaptive \ac{ppd} with natural weighting ($\sigma_{\bs r}=2.26\times 10^{-4}$), \ac{ms-clean} with natural weighting ($\sigma_{\bs r}=3.70\times 10^{-4}$), \ac{ms-clean} with Briggs weighting ($\sigma_{\bar{\bs r}}=1.77\times 10^{-4}$).} { {The lower value of the standard deviation of the residual image obtained with \ac{ppd}, compared to that obtained with naturally-weighted \ac{ms-clean} confirms the higher fidelity to data of \ac{PPD}'s estimated model image.}}}

\end{figure*}

The recovered image of adaptive \ac{PPD} is displayed in Figure \ref{fig:xmzooms}, together with the model and restored images of \ac{MS-CLEAN}. Three key regions in Cyg A are emphasised: these are the hotspots of the east and west jets (second and third column), and the inner core of the galaxy (fourth column). When inspecting the model images (rows 1 and 3 of the same figure), one can see that the maps of \ac{MS-CLEAN} present smooth extended structures since it employs non-delta functions. Though the maps remain non-physical, they are considered more reasonable when compared to the \alg{clean} algorithm \citep{hogbom74,clark1980,csclean1983}. Inspection of the restored maps of \ac{ms-clean} (rows 2 and 4) against the model image of adaptive \ac{ppd} (fifth row) shows that the latter exhibits more details significantly visible at the hotspots of Cyg A. The validity of this super-resolution is investigated in Section \ref{ssec:sr-analysis}. Furthermore, both Briggs-weighted \ac{ms-clean} and adaptive \ac{ppd} succeed in recovering a faint point source in the inner core of Cyg A, highlighted with a green circle in Figure~\ref{fig:xmzooms} (right column, rows 3-5) as opposed to naturally-weighted \ac{ms-clean} (right column, rows 1 and 2). Discussion of this radio source will follow in Section \ref{ssec:blackhole}.
Inspection of the residual images displayed in Figure \ref{fig:xres} indicates negative structures at the hotspots positions in the residual image of adaptive \ac{ppd}. These can be explained by: (i) the presence of calibration errors at those positions 
and (ii) {{employing \alg{clean} components in the self-calibration stage which can lead to biased solutions.}}
 In fact, errors in \alg{clean}-like approaches can be absorbed in the model image due to non-positivity. Imaging with the \ac{ppd} algorithmic structure where positivity is enforced on the estimate of the sky can therefore be in tension with the calibrated data. Adopting \ac{ppd} in the imaging step during the calibration phase could potentially alleviate these artefacts. 

As for the quantitative comparison of the two imaging techniques, we report the
similarity of adaptive \ac{ppd} and naturally-weighted \ac{ms-clean} $S({\tilde{\bs z}}^{\rm{\ac{ppd}}},{\tilde{\bs z}}^{\rm{\ac{ms-clean}}})= 32.23\rm{dB}$ and the similarity of adaptive \ac{ppd} and Briggs-weighted \ac{ms-clean} $S(\bar{\tilde{\bs z}}^{\rm{\ac{ppd}}},\bar{\tilde{\bs z}}^{\rm{\ac{ms-clean}}})= 32.11\rm{dB}$. These values indicate the strong agreement of the recovered low spatial frequency content with both algorithms, more precisely at the Fourier modes below the spatial band-limit of the observations.
The achieved $DR$ values with natural weighting are $6.02\times10^3 ~\rm{and}~4.26\times10^3$ for adaptive \ac{ppd} and naturally-weighted \ac{ms-clean}, respectively. This indicates the higher fidelity achieved by adaptive \ac{ppd}. On the other hand, the achieved $DR$ values with Briggs weighting are $4.2\times10^3~\rm{and~}7.76\times 10^3$ for adaptive \ac{ppd} and Briggs-weighted \ac{ms-clean}, respectively. 
Note that the latter minimises the $\ell_2$ norm of the residual image $\bar{\bs{\Phi}}^\dagger (\bar{\bs y}-\bar{\bs{\Phi}}\tilde{\bs x})$, while adaptive \ac{ppd} minimises the $\ell_2$ norm of the residual image $\bs{\Phi}^\dagger(\bs y-\bs{\Phi}\tilde{\bs x})$. Conceptually, both methods solve for different imaging problems. Therefore, the higher $DR$ achieved by Briggs-weighted \ac{ms-clean} does not necessarily imply a better performance over adaptive \ac{ppd}. On a further note, for the sake of comparison, the reported $DR$ values of adaptive \ac{ppd} are computed using the measurement operators corresponding to the two weighting schemes in \alg{wsclean}, more precisely, in the computation of the residual images.

The dynamic range on the model image of adaptive \ac{PPD} is $MDR=3.49\times10^4$ and is saturated by the DDE modelling errors. The higher value of $MDR$ compared to the $DR$ values of adaptive \ac{ppd} with both weighting schemes can be justified by two reasons. (i) In the model image of adaptive \ac{ppd}, the peak value is associated with the central nuclei; the source is super-resolved and its flux is concentrated in few pixels. However, in the adaptive \ac{ppd} \textit{restored} images (as defined in the context of \alg{clean} imaging), the source's flux is rather distributed over larger area that is of the size of the adopted \alg{clean} beam. The peak values in the {restored} images are instead associated with the hotspots. (ii) The $DR$ value, by definition, may not accurately reflect the level of the noise and errors in the restored image, consisting of the combination of the residual image and the artefacts present in the model image.

\subsubsection*{C band}
\label{ssec:cresults}
The imaged sky at a frequency of 6.678GHz is of size $3276\times 1638$ pixels, with the pixel size fixed to $0.05\arcsec$. We utilise the exact same field-of-view as in X band imaging. The spatial bandwidth of the estimated signal is $\tilde{B}_C=2.5\times B_C$, where $B_C$ is the maximum baseline. Data are split to 16 blocks with $8\times10^4$ measurements on average. We perform 60 weighted $\ell_1$ minimisation tasks using adaptive \ac{ppd}. Each minimisation task stops when the relative variation between two consecutive estimates of the sky gets below $10^{-5}$, except for the last minimisation task where the value of this stopping criterion is set to $10^{-6}$. \ac{ms-clean} imaging is performed using the weighting schemes: natural and Briggs (the robustness parameter is set to $r=-1$). Estimated model images of adaptive \ac{ppd} and \ac{ms-clean} are displayed in Figure \ref{fig:cmzooms}. Superiority of the adaptive \ac{ppd} reconstructions when compared to those of \ac{MS-CLEAN} with both weighting schemes in terms of physical surface brightness distribution and high resolution is once again confirmed. The associated residual images are displayed in Figure \ref{fig:cres}, where one can see that the adaptive \ac{ppd} residual image presents the lowest standard deviation. Furthermore, it is less structured when compared to the residual image of naturally-weighted \ac{ms-clean}.

The $DR$ values with natural weighting are $8.04\times10^3 ~\rm{and}~4.18\times10^3$ for adaptive \ac{ppd} and naturally-weighted \ac{ms-clean}, respectively. The $DR$ values with Briggs weighting are $3.45\times10^3~\rm{and~}2.77\times 10^3$ for adaptive \ac{ppd} and Briggs-weighted \ac{ms-clean}, respectively. The $MDR$ evaluated on the model image of adaptive \ac{ppd} is $1.09\times10^4$. These values indicate higher dynamic range achieved by adaptive \ac{ppd}, hence higher fidelity. In addition, we report the similarity of the model images obtained with adaptive \ac{ppd} and \ac{ms-clean} when smoothed at the instrument's resolution for the two weighting schemes. Similarity of adaptive \ac{ppd} and naturally-weighted \ac{ms-clean} is $S({\tilde{\bs z}}^{\rm{\ac{ppd}}},{\tilde{\bs z}}^{\rm{\ac{ms-clean}}})=33.51\rm{dB}$. Similarity of adaptive \ac{ppd} and Briggs-weighted \ac{ms-clean} is $S(\bar{\tilde{\bs z}}^{\rm{\ac{ppd}}},\bar{\tilde{\bs z}}^{\rm{\ac{ms-clean}}})= 32.27\rm{dB}$. Once again, these results confirm the high similarity of the low spatial frequency content of the recovered images with adaptive \ac{ppd} and \ac{ms-clean}.
\begin{figure*}
\centering
\begin{minipage}[t]{1\linewidth}
\vspace{0pt} 
\centering
\includegraphics[width=0.355\linewidth]{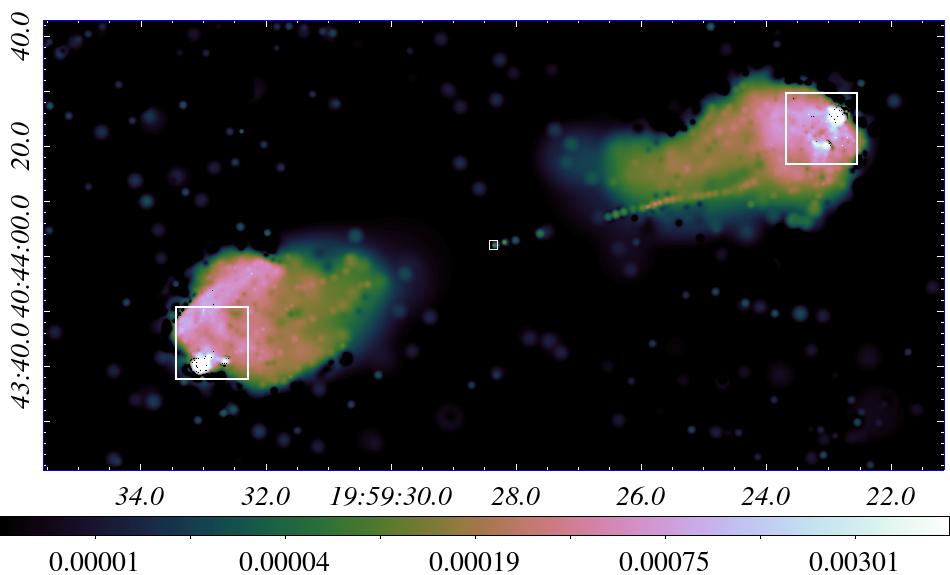}
\vspace{0pt} 
\includegraphics[width=0.201\linewidth]{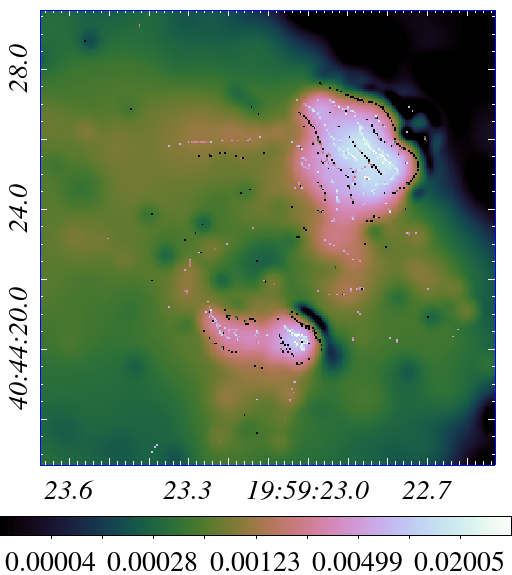}
\vspace{0pt} 
\includegraphics[width=0.201\linewidth]{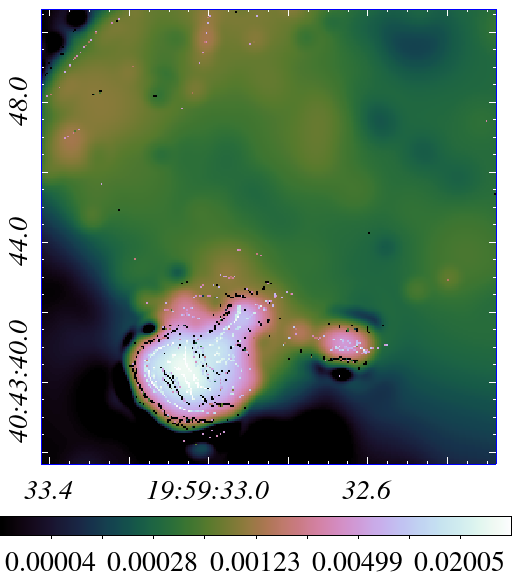}
\vspace{0pt} 
\includegraphics[width=0.201\linewidth]{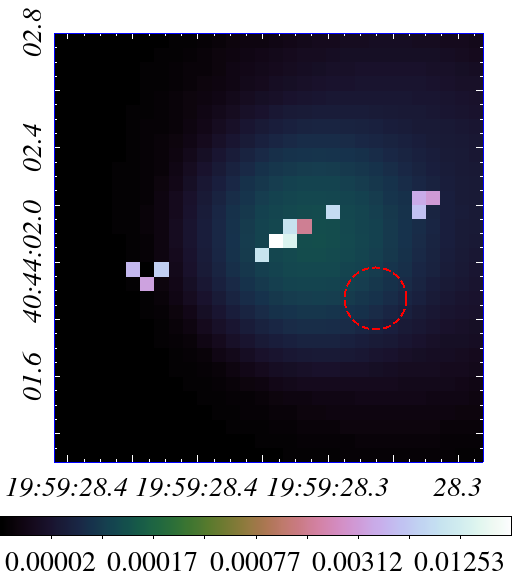}
\end{minipage}
\begin{minipage}[t]{1\linewidth}
\vspace{0pt} 
\centering
\includegraphics[width=0.355\linewidth]{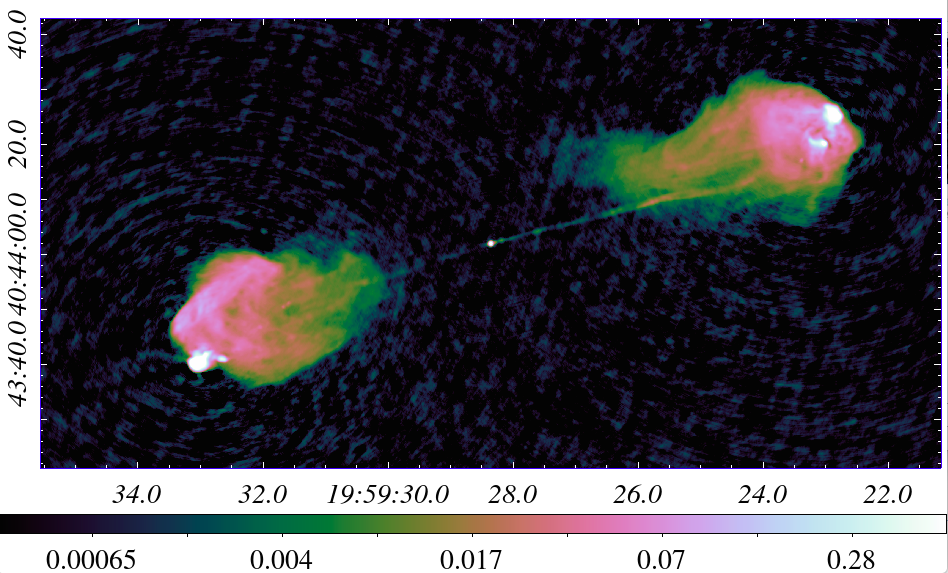}
\vspace{0pt} 
\includegraphics[width=0.201\linewidth]{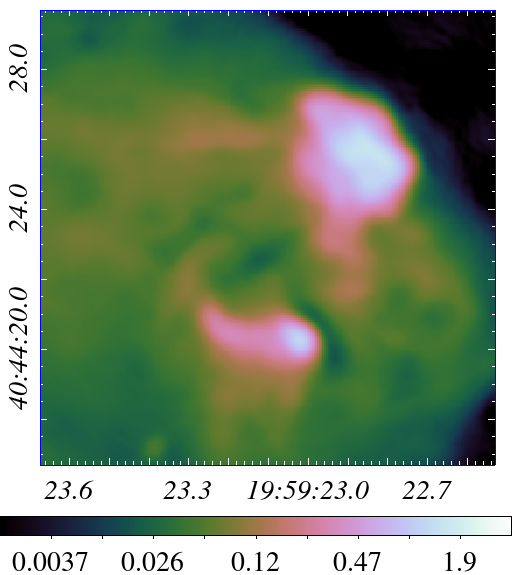}
\vspace{0pt} 
\includegraphics[width=0.201\linewidth]{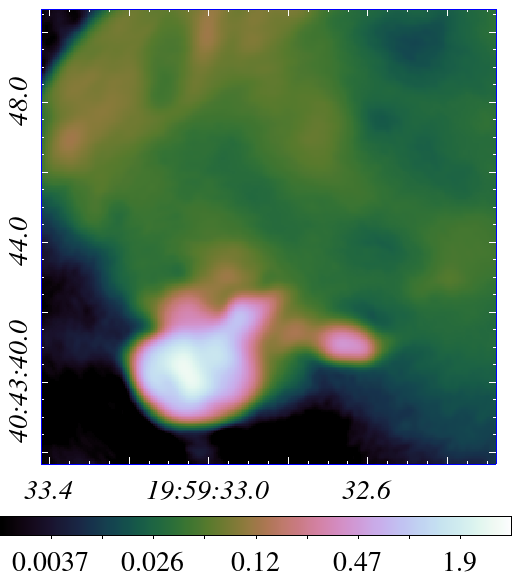}
\vspace{0pt} 
\includegraphics[width=0.201\linewidth]{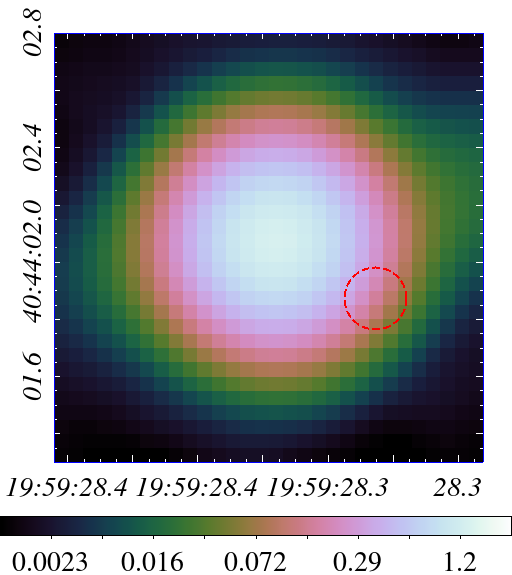}
\end{minipage}
\begin{minipage}[t]{1\linewidth}
\vspace{0pt} 
\centering
\includegraphics[width=0.355\linewidth]{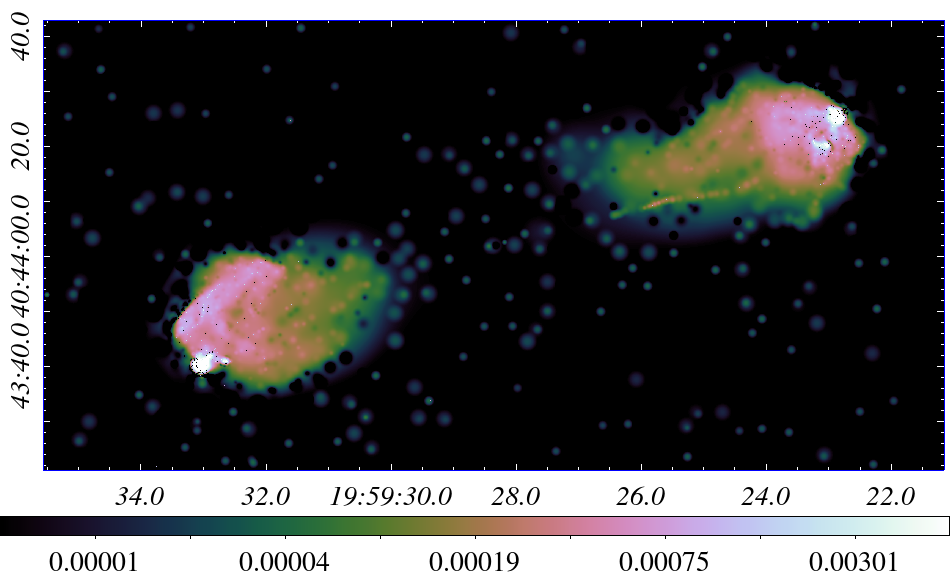}
\vspace{0pt} 
\includegraphics[width=0.201\linewidth]{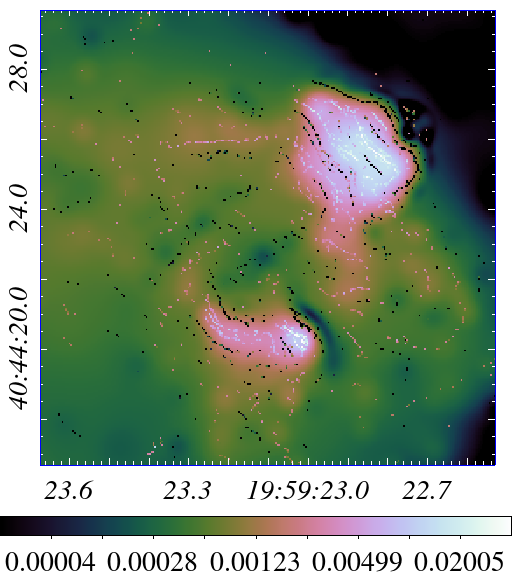}
\vspace{0pt} 
\includegraphics[width=0.201\linewidth]{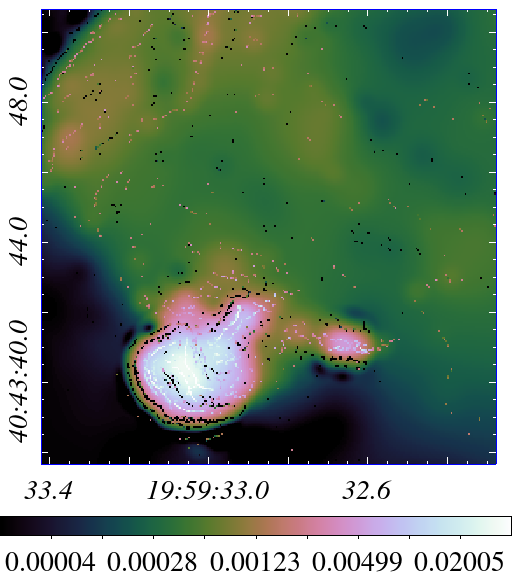}
\vspace{0pt} 
\includegraphics[width=0.201\linewidth]{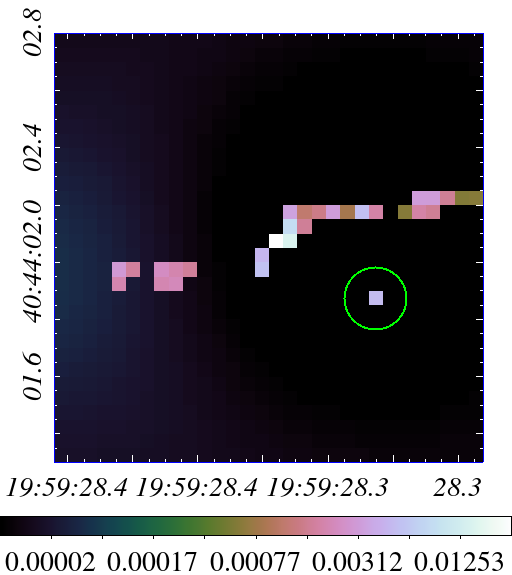}
\end{minipage}
\begin{minipage}[t]{1\linewidth}
\vspace{0pt} 
\centering
\includegraphics[width=0.355\linewidth]{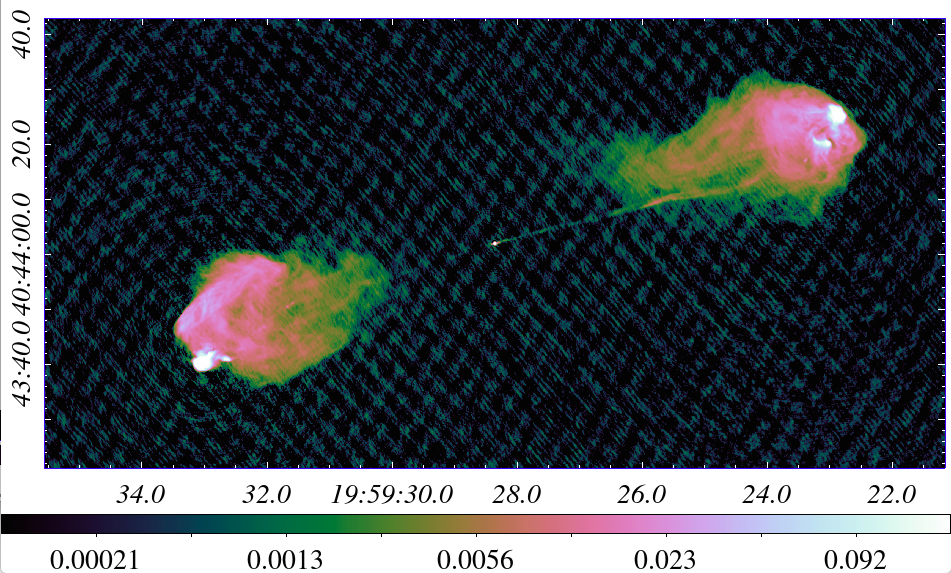}
\vspace{0pt} 
\includegraphics[width=0.201\linewidth]{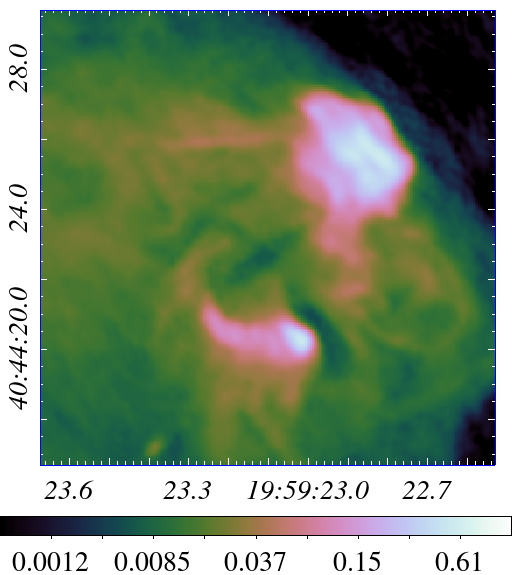}
\vspace{0pt} 
\includegraphics[width=0.201\linewidth]{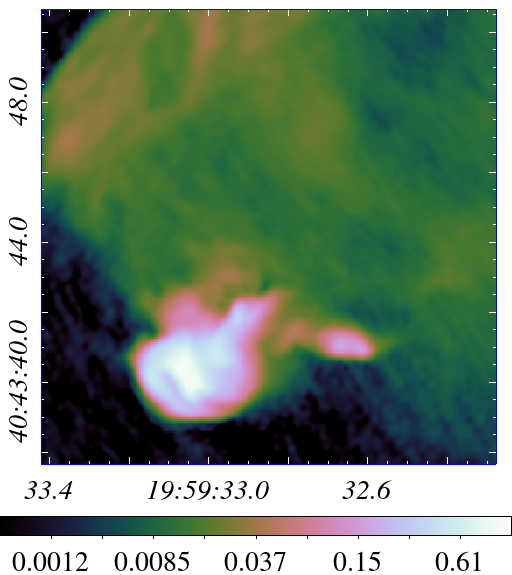}
\vspace{0pt} 
\includegraphics[width=0.201\linewidth]{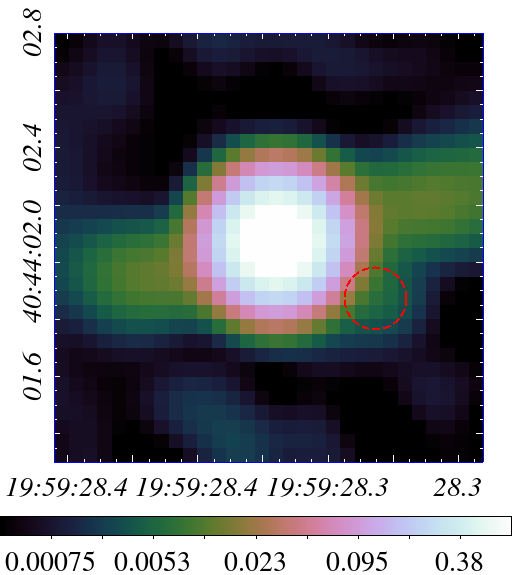}
\end{minipage}
\begin{minipage}[t]{1\linewidth}
\vspace{0pt} 
\centering
\includegraphics[width=0.355\linewidth]{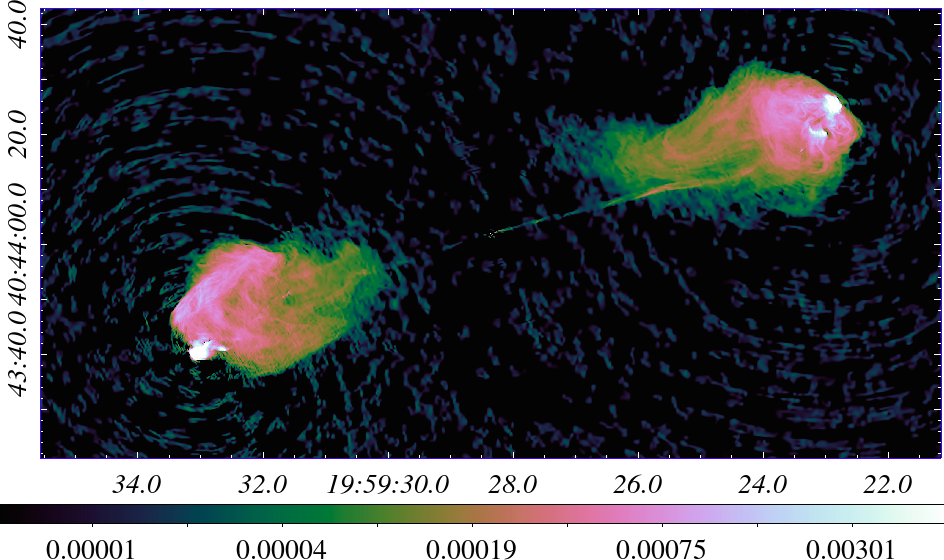}
\vspace{0pt} 
\includegraphics[width=0.201\linewidth]{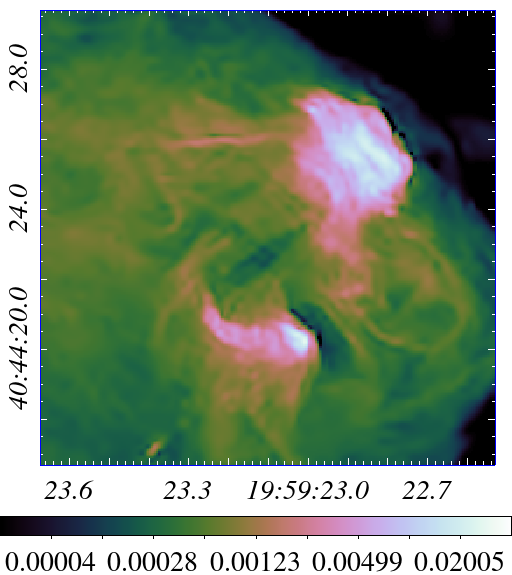}
\vspace{0pt} 
\includegraphics[width=0.201\linewidth]{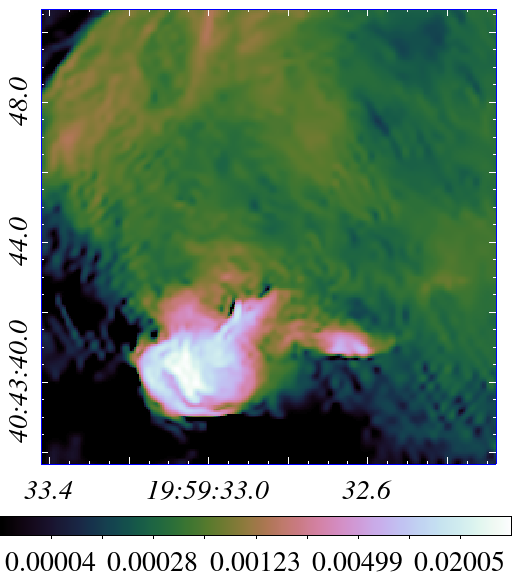}
\vspace{0pt} 
\includegraphics[width=0.201\linewidth]{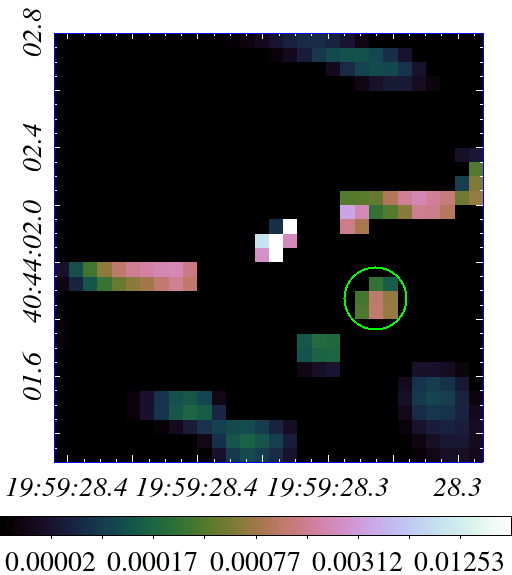}
\end{minipage}
\begin{minipage}{1\linewidth}
\caption{
\label{fig:cmzooms}
C band: recovered images at $2.5$ times the resolution of the instrument. { {From top to bottom, estimated model and restored images of naturally-weighted \ac{ms-clean} (resp. rows 1 and 2), estimated model and restored images of Briggs-weighted \ac{ms-clean} (resp. rows 3 and 4) and model image of adaptive \ac{ppd} (fifth row). The full images are displayed in $\log_{10}$ scale (first column) as well as zooms on the three brightest regions: east jet's hotspot (second column), west jet's hotspot (third column) and the inner core of the Cyg A galaxy (fourth column). The zoomed regions are highlighted with white boxes in the model image of naturally-weighted \ac{ms-clean} (top row, left column). The surface brightness of the restored image obtained with naturally-weighted \ac{ms-clean} (second row) is in units of ${\rm{Jy}}/{\rm{beam}}$, the naturally-weighted beam is of size $0.45\arcsec \times 0.45\arcsec$ and its flux is $93.65\rm{Jy}$.  The surface brightness of the restored image obtained with Briggs-weighted \ac{ms-clean} (fourth row) is also in units of ${\rm{Jy}}/{\rm{beam}}$, the Briggs-weighted beam is of size $0.25\arcsec \times 0.25\arcsec$ and its flux is $30.44{\rm{Jy}}$.
The surface brightness of the model images (rows 1, 3 and 5) is in units of ${\rm{Jy}}/{\rm{pixel}}$, the pixel size being $0.05\arcsec$ in both directions.}}
}
\end{minipage}
\end{figure*}

\begin{figure*}
\vspace{0pt}
\includegraphics[width=0.3\linewidth]{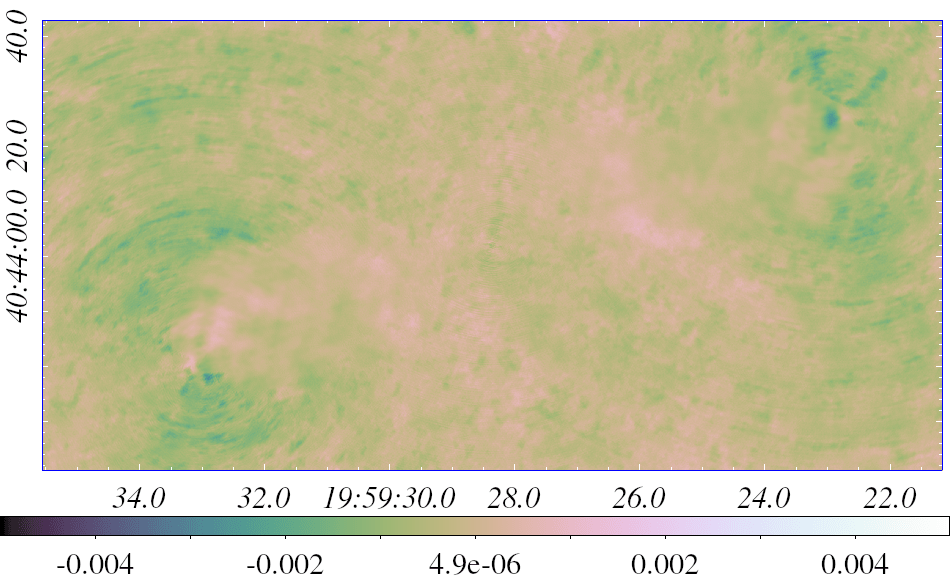}
\vspace{0pt} 
\includegraphics[width=0.3\linewidth]{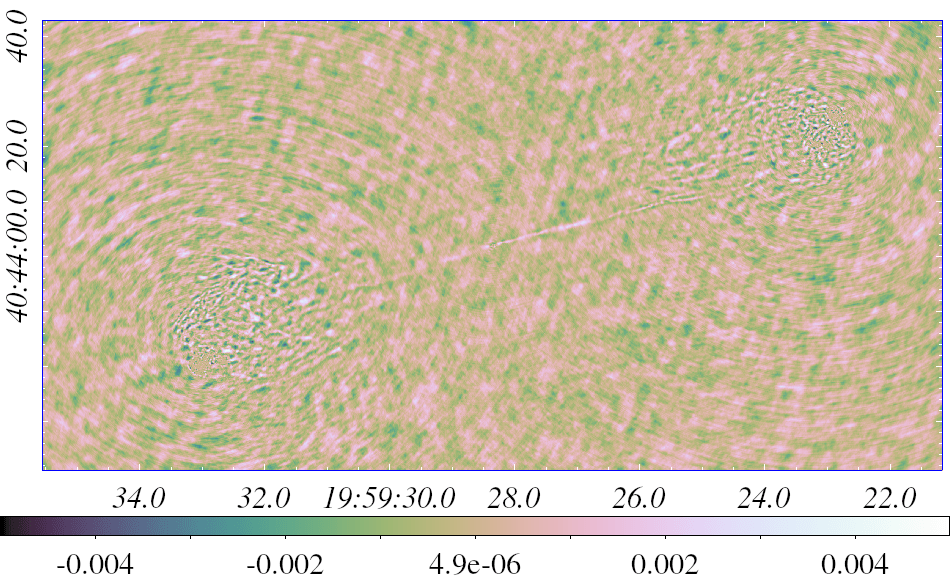}
\vspace{0pt} 
\includegraphics[width=0.3\linewidth]{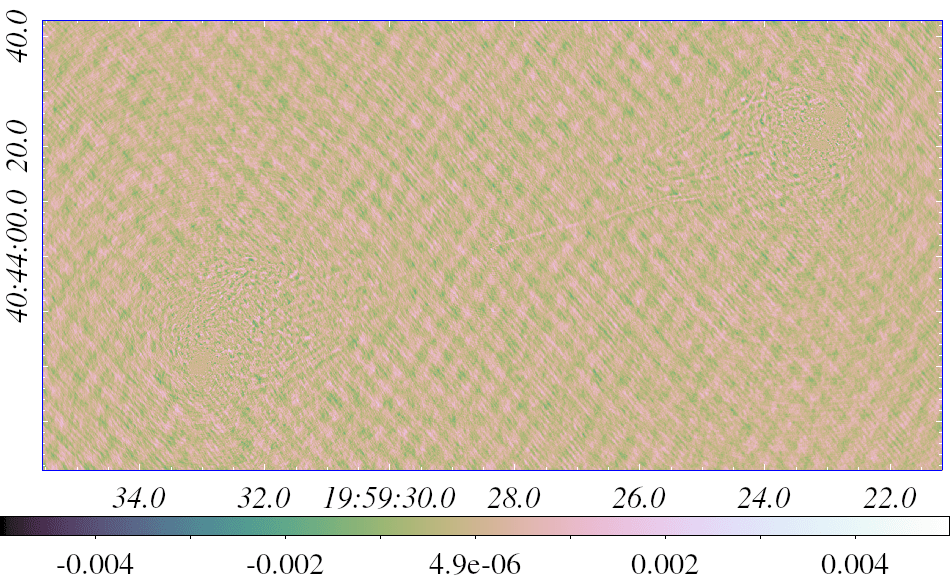}
\caption{
\label{fig:cres} 
{C band: residual images. From left to right: images of adaptive \ac{ppd} with natural weighting ($\sigma_{\bs r}=3.65\times 10^{-4}$), \ac{ms-clean} with natural weighting ($\sigma_{\bs r}=7.01\times 10^{-4}$), \ac{ms-clean} with Briggs weighting ($\sigma_{\bar{\bs r}}=4.60\times 10^{-4}$).} { {The lower value of the standard deviation of the residual image obtained with \ac{ppd}, compared to that obtained with naturally-weighted \ac{ms-clean} confirms the higher fidelity to data of \ac{PPD}'s estimated model image. }}}
\end{figure*}

\section{Super-resolution of Cyg A}
\label{ssec:sr}
Both recovered images of Cyg A at bands X and C  with adaptive \ac{ppd} exhibit high resolution features when compared to the restored maps of \ac{ms-clean}. In this section, we analyse the super-resolution achieved with adaptive \ac{PPD} by referring to higher resolution observations of Cyg A. We also confirm the detection of a secondary black hole in the inner core of Cyg A, reported in \cite{perley2017} and super-resolved in the maps produced by adaptive \ac{ppd}.

\subsection{Analysis of the super-resolution with adaptive \ac{ppd} }
\label{ssec:sr-analysis}
To judge the veracity of the super-resolution capabilities of adaptive \ac{ppd}, we compare its results to maps of Cyg A obtained from higher-frequency observations, where the nominal resolution is naturally higher, hence super-resolution is not required. More precisely, we choose VLA observations at { {Ku band ($12-18~\rm{GHz}$)}}, where the nominal resolution is two to three times that of the X and C band data. Their associated maps are restored images obtained with the Cotton-Schwab \alg{clean} \citep{csclean1983}. Since the emission mechanism in Cyg A is known to be synchrotron, the radiation spectrum is very broad, spanning over orders of magnitudes \citep{carilli91}. No sharp spectral features are expected within a frequency band of the same order of magnitude. Therefore, the choice of Ku band images as a reference is relevant. 

For the imaged data at X band, we utilise a restored map of Cyg A at a frequency of 17.324GHz as a reference. These data present a maximum baseline $B_1^{\rm{ref}}$, that is $B_1^{\rm{ref}}\approx2\times {B_X}$, and the spatial bandwidth of its imaged map is $\tilde{B}_{1}^{\rm{ref}}=4\times B_X$. The X band data, initially imaged at the spatial bandwidth $2.5\times B_X$, are further imaged at the exact same spatial bandwidth as the reference image $\tilde{B}_{1}^{\rm{ref}}$ using both Briggs-weighted \ac{ms-clean} and adaptive \ac{ppd}. In Figure \ref{fig:xsr}, zooms on the hotspots are displayed for the reference image, the model image of adaptive \ac{ppd} and the restored image of Briggs-weighted \ac{ms-clean}. All three images are characterised with a pixel size $\delta l= 0.025\arcsec$. { { The surface brightness of \ac{PPD} model images are in units of Jy/pixel and that of \ac{MS-clean} and the reference maps are in units of Jy/beam. Note that the displayed integrated flux is preserved on the three maps.}} The inspection of the zooms on both hotspots indicates the high similarity of the recovered hotspots in the model image of adaptive \ac{ppd} and the \alg{clean} restored image of Cyg A at the frequency 17.324GHz. A further examination of adaptive \ac{ppd}'s consistency with respect to the choice of the imaging resolution is examined through the image recovery at three different resolutions. These correspond to the pixel sizes $\delta l \in \{0.08\arcsec$, $0.04\arcsec, 0.025\arcsec \}$, shown in Figure \ref{fig:xsr} as embedded animations\footnote{The animation is only supported when the PDF file is opened using Adobe Acrobat Reader, https://get.adobe.com/reader/} cycling through the imaged hotspots at the different resolutions. The high resolution features are consistent over the different resolutions with a noticeable improvement when increasing the imaged spatial bandwidth, in particular in terms of the pixelisation at the edges of the hotspots' brightest structures. Thus, super-resolved maps up to four times the nominal resolution can be obtained with no apparent degradation of the imaging quality despite the increase of the number of the unknowns in the imaging problem. {{Note that when running \ac{ms-clean} on a resolution larger than the nominal one, naturally a super-resolved model image is obtained, where the negative components are less prominent. However, as explained earlier, the super-resolved structures of the \ac{ms-clean} model image remain non-physical. Moreover, the resolution of the corresponding restored image is limited by the shape of the \alg{clean} beam, dictated by the effective Fourier sampling. This effect is illustrated in Figure \ref{fig:xsr}, where one can see that (i) the hotspots recovered in the restored image of Briggs-weighted \ac{ms-clean} are smooth when compared against the reconstruction of adaptive \ac{ppd}, (ii) no super-resolution is noticed when examining the hotspots at the three different resolutions.}}

Similar investigation is carried out for the obtained images from the data at C band. These are cross-checked with a restored image of Cyg A at a frequency of 14.252GHz. The maximum baseline of the observations is $B_2^{\rm{ref}}\approx 2.13\times B_C$ and the imaged map's spatial bandwidth is $\tilde{B}_{2}^{\rm{ref}}\approx 3.5\times B_C$ corresponding to a pixel size $\delta l=0.035\arcsec$. We perform imaging with adaptive \ac{ppd} and Briggs-weighted \ac{ms-clean} at the same spatial bandwidth $\tilde{B}_{2}^{\rm{ref}}$ (\emph{i.e.} same resolution). Zooms on the hotspots of the reference map, the adaptive \ac{PPD} model image and the Briggs-weighted \ac{ms-clean} restored image are shown in Figure \ref{fig:csr}. The surface brightness of \ac{PPD} model images are in units of Jy/pixel and that of \ac{MS-clean} and the reference maps are in units of Jy/beam. Once again, when inspecting visually the hotspots, it is clear that the obtained details with adaptive \ac{ppd} are physical. Super-resolution recovery of adaptive \ac{ppd} is again confirmed. Moreover, inspecting the obtained maps with adaptive \ac{ppd} at the different resolutions, characterised with pixel sizes $\delta l \in \{0.08\arcsec,0.05\arcsec,0.035\arcsec \}$, demonstrates the consistency of the high resolution features of the algorithm. These images are shown in Figure \ref{fig:csr} as embedded animations cycling through the imaged hotspots at the different resolutions. 

These super-resolved maps of adaptive \ac{ppd} are the result of the SARA priors enforced on the estimate of the sky, which are the positivity and the re-weighted average sparsity in a redundant dictionary. In particular, positivity seems to have a significant impact on  the spatial frequencies above the maximum baseline of the observations. For a better understanding of this prior's contribution, we note that the reconstructed image of the radio sky is a sampling of a spatially band-limited version of the true sky. Moreover, if exact, it can be expressed as the sampled convolution of the true positive sky with a \textit{sinc} function. Thus, conceptually, it can take negative values. Therefore, formally, the positivity constraint comes in tension with the nature of the true samples of the spatially band-limited sky. Yet, this tension can be alleviated by choosing to image the radio sky at a spatial bandwidth significantly larger than the maximum baseline of the observations. In this case, positivity simply acts as a strong prior for the Fourier modes beyond the maximum baseline of the observations, \emph{i.e.} for super-resolution. As shown above, super-resolution obtained with adaptive \ac{ppd} is validated. These findings confirm that the estimated model images using convex optimisation techniques are highly reliable, thus no post-processing such as introducing a blur reflecting the instrument's resolution, is required. On a further note, the $w$-modulation, originating from the third dimension of the baseline, has recently been shown to yield super-resolution \citep{Dabbech17}. Yet, this is not the case here, as the probed field-of-view is narrow. The $w$-modulation, being negligible, is not considered in the imaging problem.
\begin{figure*}
\begin{minipage}{1\linewidth}
\centering
\includegraphics[height=0.35\linewidth]{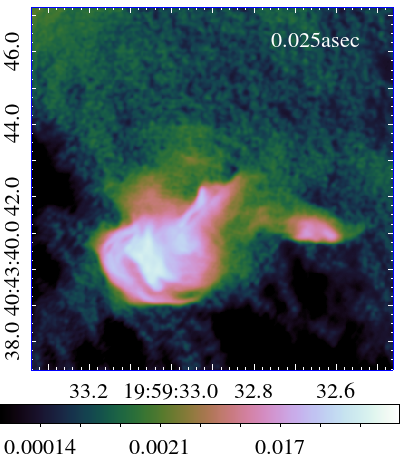}
{  \animategraphics[height=0.35\linewidth, controls]{1}{figs/gif2/xsr_ppd_}{0}{3}
}
{  \animategraphics[height=0.35\linewidth, loop, controls]{1}{figs/gif2/xsr_ws_}{0}{3}
}
\end{minipage}
\begin{minipage}{1\linewidth}
\centering
\includegraphics[height=0.35\linewidth]{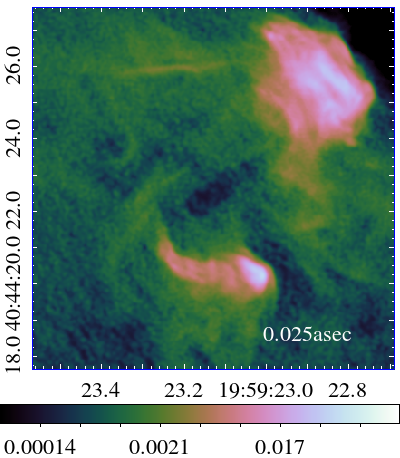}
{\animategraphics[height=0.35\linewidth, loop, controls]{1}{figs/gif2/xsr2_ppd_}{0}{3}
}
{\animategraphics[height=0.35\linewidth, loop, controls]{1}{figs/gif2/xsr2_ws_}{0}{3}
}
\end{minipage}
\caption{X band: zooms on the hotspots in Cyg A displayed in $\log_{10}$ scale. Top: east hotspot, bottom: west hotspot. From left to right: the reference map at 17.324GHz obtained with \alg{clean}, the estimated model image of adaptive \ac{ppd} and the restored image of Briggs-weighted \ac{ms-clean} from the data at X band (8.422GHz). All images have the same pixel size $\delta l=0.025\arcsec$. The figure also contains embedded
animation of the hotspots imaged with adaptive \ac{PPD} and Briggs-weighted \ac{ms-clean} at three different resolutions corresponding to $\delta l \in \{0.08\arcsec,0.04\arcsec,0.025\arcsec\}$. The surface brightness of Briggs-weighted \ac{MS-clean} and the reference map are in units of Jy/beam. The surface brightness of \ac{PPD}'s model images are in units of Jy/pixel. Note that, for each resolution, the unit of the surface brightness is different as it is a function of the pixel. The displayed integrated flux is preserved on all the maps. One can notice improved details with increased resolution, in particular at the edges of the brightest structures of the hotspots. 
The animations are only supported when the PDF file is
opened using Adobe Acrobat Reader.}
	\label{fig:xsr}
\end{figure*}

\begin{figure*}
\begin{minipage}{1\linewidth}
\centering
\includegraphics[height=0.35\linewidth]{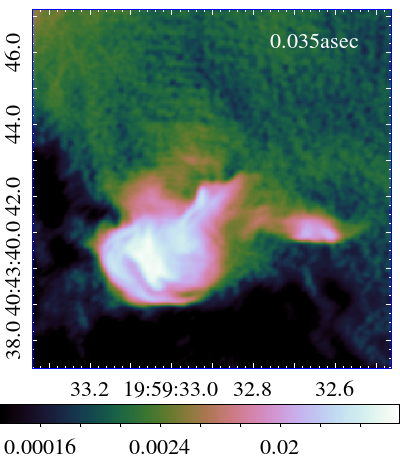}
{  \animategraphics[height=0.35\linewidth, loop, controls]{1}{figs/gif2/csr_ppd_}{0}{3}
}
{  \animategraphics[height=0.35\linewidth, loop, controls]{1}{figs/gif2/csr_ws_}{0}{3}
}
\end{minipage}
\begin{minipage}{1\linewidth}
\centering
\includegraphics[height=0.35\linewidth]{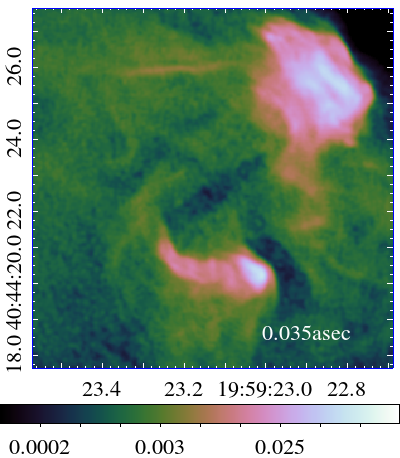}
{  \animategraphics[height=0.35\linewidth, loop, controls]{1}{figs/gif2/csr2_ppd_}{0}{3}
}
{  \animategraphics[height=0.35\linewidth, loop, controls]{1}{figs/gif2/csr2_ws_}{0}{3}
}
\end{minipage}

\caption{C band: zooms on the hotspots in Cyg A displayed in $\log_{10}$ scale. Top: east hotspot, bottom: west hotspot. From left to right: the reference map at 14.252GHz obtained with \alg{clean}, the estimated model image of adaptive \ac{ppd} and the restored image of Briggs-weighted \ac{ms-clean} from the data at C band (6.678GHz). All images have the same pixel size $\delta l=0.035\arcsec$.{ {The surface brightness of Briggs-weighted \ac{MS-clean} and the reference map are in units of Jy/beam. The surface brightness of \ac{PPD}'s model images are in units of Jy/pixel. Note that, for each resolution, the unit of the surface brightness is different as it is a function of the pixel. The displayed integrated flux is preserved on all the maps}}.  The figure also contains embedded
animations of the hotspots imaged with adaptive \ac{PPD} and Briggs-weighted \ac{Ms-clean} at three different resolutions, corresponding to $\delta l \in \{0.08\arcsec,0.05\arcsec,0.035\arcsec \}$. One can notice improved details with increased resolution, in particular at the edges of the brightest structures of the hotspots.
The animations are only supported when the PDF file is
opened using Adobe Acrobat Reader. }
	\label{fig:csr}
\end{figure*}

\subsection{The story of a secondary black hole (candidate)}
\label{ssec:blackhole}
\cite{perley2017} report the serendipitous discovery of a luminous radio transient in the inner core of Cyg A, just 460 pc offset from the super massive black hole in the galaxy. The transient, dubbed Cyg A-2, is well detected at 8.5GHz using \alg{clean}, and is interpreted as a secondary black hole. 
We confirm the findings of \cite{perley2017} when imaging Cyg A from observations at a frequency of 8.422GHz, at 2.5 times the nominal resolution (corresponding to a pixel size of $\delta l=0.04\arcsec$). In Figure \ref{fig:xmzooms}, fourth column, zooms on the inner core of Cyg A are displayed. { {The source's location is highlighted with a circle whose center is at the position given by $\rm{RA}=19\rm{h}~59\rm{mn}~28.322\rm{s}$ ($J2000$) and $\rm{DEC}=+40^{\circ}~44\arcmin~1.89\arcsec$ and radius of size $0.1\arcsec$. The source is highlighted with a green circle when detected and a red dashed circle otherwise.}}  One can see that Cyg A-2 is well detected with adaptive \ac{ppd} and Briggs-weighted  \ac{ms-clean}, whereas naturally-weighted \ac{ms-clean} fails to do so. Though visible, the source is blurred in the restored image of Briggs-weighted \ac{ms-clean}. The flux of Cyg A-2, calculated directly from the model images (over the highlighted physical region), is about $5\rm{mJy}$ with adaptive \ac{ppd} and $4.9\rm{mJy}$ with Briggs-weighted \ac{ms-clean}. {  The source is also well detected when imaged at 4 times the nominal resolution (corresponding to a pixel size $\delta l=0.025\arcsec$) as shown in Figure \ref{fig:xsrz1}, left panel, where its angular scale is preserved and its estimated flux is about $4.3\rm{mJy}$.}

\begin{figure}
\begin{minipage}{0.47\linewidth}
\centering
{  \animategraphics[label=axsr,height=1\linewidth, loop, controls]{1}{figs/gif2/axsr_z1}{0}{2}

}
\end{minipage}
\begin{minipage}{0.47\linewidth}
\centering
{  \animategraphics[loop,label=acsr,height=1\linewidth, controls]{1}{figs/gif2/acsr_z1}{0}{2}
}
\end{minipage}
\caption{{  {Cyg A-2 displayed in $\log_{10}$ scale. Left, X band observations imaged with \ac{ppd} at 4 times the nominal resolution ($\delta l=0.025\arcsec$).  Right, C band observations imaged with \ac{ppd} at 3.5 times the nominal resolution ($\delta l=0.035\arcsec$). Cyg A-2 is highlighted with a green circle. The source's angular scales from the maps at 2.5 times the nominal resolutions are highlighted with green dashed circles. These maps are embedded as animations in the figure. The animations are only supported when the PDF file is
opened using Adobe Acrobat Reader.}}}
	\label{fig:xsrz1}
\end{figure}

More interestingly, the source is highly resolved when imaged with adaptive \ac{ppd} from the observations at a frequency of 6.678GHz at 2.5 times the nominal resolution (corresponding to a pixel size $\delta l=0.05\arcsec$), as shown in Figure \ref{fig:cmzooms}, fourth column. { {The source's location is highlighted with a circle whose center is at the position given by $\rm{RA}=19\rm{h}~59\rm{mn}~28.324\rm{s}$ ($J2000$) and $\rm{DEC}=+40^{\circ}~44\arcmin~1.88\arcsec$ and radius of size $0.11\arcsec$. The source is highlighted with a green circle when detected and a red dashed circle otherwise. }}Although detected in the model image of \ac{ms-clean} with Briggs weighting at a single pixel, in the restored image, the source is completely buried within the beam of the primary nuclei of Cyg A (see Figure \ref{fig:cmzooms}, fourth column, fourth row). As for \ac{ms-clean} with natural weighting, here again it fails completely to detect the radio transient. { {The estimated flux of Cyg A-2 is about $4.6\rm{ mJy}$ with both adaptive \ac{ppd} and Briggs-weighted \ac{ms-clean} (over the highlighted physical region).}} These results are in agreement with the findings of \cite{perley2017}. {  The source is also well detected when imaged at 3.5 times the nominal resolution (corresponding to a pixel size $\delta l=0.035\arcsec$) as shown in Figure \ref{fig:xsrz1}, right panel, where it is further resolved and its estimated flux is about $3.4\rm{mJy}$. We note the presence of a tail-like structure associated with Cyg A-2 in the \ac{PPD} image. However, given the faintness of the structure, and the fact that it is not detected in the X band image, we cannot confidently say whether this faint structure is real, or is a DDE-induced imaging artefact.}

\section{Conclusions}
\label{sec:cend}
In this paper, we developed an adaptive version of the convex \ac{ppd} algorithmic structure solving the SARA minimisation problem for radio interferometric imaging in the presence of unknown noise and calibration errors. The algorithm achieves high resolution high fidelity imaging of Cyg A from VLA observations. 
Imaging results confirm the superior quality of the proposed algorithmic structure to standard \alg{clean}-based techniques. The veracity of the achieved super-resolved reconstructions of Cyg A at X and C bands is verified through higher-resolution VLA observations of the radio galaxy at Ku band. These results confirm the reliability of the reconstructed images with the advanced convex optimisation algorithms as accurate representations of the radio sky. Our \alg{matlab} code is available online on GitHub,
{http://
basp-group.github.io/pd-and-admm-for-ri/}. Interestingly, the recent discovery of a radio transient in the inner core of Cyg A, revealed at X band, is further confirmed at C band when imaging with adaptive \ac{PPD}, though the latter observations are characterised with a lower nominal resolution. The radio source is very well detected on the  adaptive \ac{ppd} model images at both frequencies and is super-resolved when compared against the restored images obtained with \ac{ms-clean}.

\section*{Acknowledgements}
The National Radio Astronomy Observatory is a facility of the National Science Foundation operated under cooperative agreement by Associated Universities, Inc. The research of AD, AO, AA and YW is supported by the UK Engineering and Physical Sciences Research Council (EPSRC, grants EP/M008843/1, EP/M011089/1). The research of OS is supported by the South African Research Chairs Initiative of the Department of Science and Technology and National Research Foundation.




\bibliographystyle{mnras}
\bibliography{biblio}



\appendix
\section{Overview of the parameters specific to Adaptive \ac{PPD}}
\label{apx:a}
An overview of the variables and parameters involved in the adjustment of the $\ell_2$ bounds on the data fidelity terms is presented in Tables \ref{tab:param-l2-adjust} and \ref{tab:var-l2-adjust}, respectively.
\begin{table}
	\bc
 	\caption{Overview of the variables employed in the adaptive procedure incorporated in Algorithm~\ref{alg-primal-dual}.}
 	\label{tab:var-l2-adjust}
 	\centering
	\small
 	\begin{tabular}{p{1.4cm}p{6cm}}
	\hline
	$\mu_j^{(t)}$ & $\ell_2$ norm of the residual corresponding to the data block $j$ at iteration $t$. \\
	$\epsilon_j^{(t-1)}$ & $\ell_2$ bound on the data block $j$ imposed at iteration $t$.\\
	$p_j^{(t-1)}$ & iteration index of the previous update of the $\ell_2$ bound of the data block ${j}$.\\
	$\sigma^{(t-1)}$ & characterising the relative variation between two consecutive estimates of the solution at iteration $t-1$.\\
 
	\hline 
 	\end{tabular}%
	\ec
\end{table}
\begin{table}
	\bc
 	\caption{Overview of the parameters involved in the adaptive procedure incorporated in Algorithm~\ref{alg-primal-dual}.}
 	\label{tab:param-l2-adjust}
 	\centering
	\small
 	\begin{tabular}{p{1.4cm}p{6cm}}
	\hline
	$\gamma_1 \in ] 0 ~1[ $ & configurable; the bound on the relative variation between two consecutive estimates of the solution. For the tests herein $\gamma_1$ is set to $10^{-4}$.\\
	$\gamma_2 \in ] 0 ~1[ $ & configurable; the tolerance on the relative difference between the current estimate of the bound imposed on the data block $j$ and the $\ell_2$ norm of its associated residual. For the tests herein $\gamma_2$ is set to $10^{-3}$.\\
	$\gamma_3 \in ] 0 ~1[ $ & configurable, characterising the increment of the $\ell_2$ bound with respect to the $\ell_2$ norm of the current residual. For the tests herein $\gamma_3$ is set to $0.618$ \\
	$P$ & configurable; corresponds to the minimum number of iterations between consecutive updates on each $\ell_2$ bound. For the tests herein $P$ is set to 100.\\
	\hline 
 	\end{tabular}%
	\ec
\end{table}


\bsp	
\label{lastpage}
\end{document}